\documentclass[12pt,onecolumn,draftcls]{IEEEtran}
\usepackage{graphicx}
\usepackage{amsmath,amssymb,epsfig}
\usepackage{subfigure}
\usepackage[sort&compress,square,numbers]{natbib}
\usepackage{blkarray}
\usepackage{arydshln}
\usepackage{enumerate}
\usepackage{color}

\newtheorem{theorem}{Theorem}
\newtheorem{proposition}[theorem]{Proposition}
\newtheorem{lemma}[theorem]{Lemma}
\newtheorem{corollary}[theorem]{Corollary}
\newtheorem{assumption}{Assumption}
\newtheorem{claim}[theorem]{Claim}
\newtheorem{observation}[theorem]{Observation}

\newtheorem{definition}{Definition}

\newcommand{\beq}{\begin{equation}}
\newcommand{\eeq}{\end{equation}}
\newcommand{\bea}{\begin{array}}
\newcommand{\ena}{\end{array}}
\newcommand{\bds}{\begin {itemize}}
\newcommand{\eds}{\end {itemize}}
\newcommand{\bdf}{\begin{definition}}
\newcommand{\blm}{\begin{lemma}}
\newcommand{\edf}{\end{definition}}
\newcommand{\elm}{\end{lemma}}
\newcommand{\bthm}{\begin{theorem}}
\newcommand{\ethm}{\end{theorem}}
\newcommand{\bprp}{\begin{prop}}
\newcommand{\eprp}{\end{prop}}
\newcommand{\bcl}{\begin{claim}}
\newcommand{\ecl}{\end{claim}}
\newcommand{\bcr}{\begin{coro}}
\newcommand{\ecr}{\end{coro}}
\newcommand{\bquest}{\begin{question}}
\newcommand{\equest}{\end{question}}

\def\bm#1{\mbox{\boldmath $#1$}}

\newcommand{\bSigma}{\mbox{$\bm \Sigma$}}

\newcommand{\avec}{{\bf{a}}}

\newcommand{\yvec}{{\bf{y}}}
\newcommand{\uvec}{{\bf{u}}}
\newcommand{\wvec}{{\bf{w}}}
\newcommand{\xvec}{{\bf{x}}}
\newcommand{\zvec}{{\bf{z}}}

\newcommand{\rvec}{{\bf{r}}}

\newcommand{\svec}{{\bf{s}}}
\newcommand{\vvec}{{\bf{v}}}

\newcommand{\Amat}{{\bf{A}}}
\newcommand{\Bmat}{{\bf{B}}}

\newcommand{\Gmat}{{\bf{G}}}
\newcommand{\Hmat}{{\bf{H}}}

\newcommand{\Imat}{{\bf{I}}}

\newcommand{\Pmat}{{\bf{P}}}

\newcommand{\Tmat}{{\bf{T}}}

\newcommand{\Rmat}{{\bf{R}}}
\newcommand{\Umat}{{\bf{U}}}
\newcommand{\Vmat}{{\bf{V}}}
\newcommand{\Wmat}{{\bf{W}}}

\newcommand{\E}{{\rm{E}}}

\newcommand{\Tr}{{\rm Tr}}

\newcommand{\real}{{\mathbb{R}}}
\newcommand{\comp}{{\mathbb{C}}}
\newcommand{\nat}{{\mathbb{N}}}

\newcommand{\0}{{\mathbf {0}}}

\newcommand{\define}{\stackrel{\triangle}{=}}

\def\bLambda{{\mbox{\boldmath $\Lambda$}}}

\def\gammavec{{\mbox{\boldmath $\gamma$}}}

\def\phivec{{\mbox{\boldmath $\phi$}}}

\newcommand{\be}{\begin{equation}}
\newcommand{\ee}{\end{equation}}
\newcommand{\beqna}{\begin{eqnarray}}
\newcommand{\eeqna}{\end{eqnarray}}

\title{ Blind Null-Space Learning for MIMO Underlay Cognitive Radio Networks }

\author{Yair Noam and Andrea J. Goldsmith, \IEEEmembership{Fellow, IEEE}\thanks{This work is supported by the ONR under grant N000140910072P00006,
the AFOSR under grant FA9550-08-1-0480, and the DTRA under grant HDTRA1-08-1-0010. The authors are with the Dept. of Electrical Engineering, Stanford University, Stanford CA, 940305.}}

\begin{document}

\maketitle

\begin{abstract}
This paper proposes a blind technique for MIMO  cognitive radio Secondary Users (SU) to transmit in the same band   simultaneously  with a Primary User (PU) under a maximum interference constraint.   In the proposed technique, the SU is  able to  meet the interference constraint   of  the   PU without explicitly estimating the interference channel matrix to the PU and without burdening the PU with any interaction   with the  SU.
  The only condition required of the PU is that for a short time interval it     uses a power control scheme such that  its transmitted power is a monotonic function of the interference inflicted by the   SU.  During this time interval, the SU  iteratively  modifies the spatial orientation of its   transmitted signal and measures  the effect of this modification on  the PU's total transmit power.
The entire process is based  on energy measurements which is very  desirable from an implementation point of view.  \end{abstract}

\section{Introduction}

The emergence of Multiple Input Multiple Output (MIMO) communication opens new directions  and possibilities for  Cognitive Radio (CR) networks
\citep{ZhangExploiting2008,
ScutariCognitive2008,ScutariMIMO2010}. In particular, in underlay CR networks, MIMO technology enables the SU to transmit a significant amount of  power   simultaneously in the same band with the  PU  without interfering with him if the SU utilizes separate spatial dimensions  than the PU. This spatial separation requires that the  interference channel from the SU to the PU be known to the SU. Thus, acquiring  this knowledge, or   operating without it, is a major issue of active research
\citep{HuangDecentralized2011,
ZhangRobust2009,ZhangOptimal2011,
ChenInterference2011,YiNullspace}.
 We consider MIMO primary and secondary systems defined as follows: we assume  a flat-fading   MIMO  channel with  one primary and multiple  SUs.   Let  ${\bf H}_{1j}\in$  ${\mathbb C}^{n_{1} \times n_{j}}$   be  the  channel matrix between  user $j$'s  transmitter   and the PU's  receiver.  In the underlay CR paradigm,   SUs   are    constrained not to exceed a maximum interference level at  the PU, i.e.
\beq
\label{InterferenceConstraint}
\|\Hmat_{1j}\xvec_{j}(t)\|^{2}\leq \eta, ~\forall j\neq 1,
\eeq
where \(\xvec_{j}\) is SU \(j\)'s (\(j>1\)) transmit signal and \(\eta>0\) is the maximum interference constraint. If      $\eta=0$, the   SUs are strictly constrained  to transmit only within the null space of the matrix $\Hmat_{1j}$.

The optimal power allocation for   the  case of a single SU who  knows the  matrix
 $\Hmat_{21}$
 in addition to its own Channel State Information (CSI) was derived by Zahng and Liang
 \citep{ZhangExploiting2008}.
 For the case of multiple SUs, Scutari at al.
 \citep{ScutariMIMO2010}
formulated a  competitive game between the secondary users. Assuming that the interference matrix to the PU is known by each SU, they derived conditions  for the existence and uniqueness of a Nash Equilibrium point to the game. Zhang et al.
 \citep{ZhangRobust2009}
 were the first to take into consideration the fact that the  interference matrix
$\Hmat_{12}$
may not be  perfectly known  (but is partially known) to the SU. They  proposed Robust Beamforming  to assure compliance with the interference constraint of the PU while maximizing the SU's
 throughput. Other  work  for the case of an unknown interference channel with known probability distribution is due to
Zhang and So \citep{ZhangOptimal2011}, who optimized the SU throughput under a constraint on the  maximum probability that the interference to the PU is above a threshold.

The underlay concept  of CR in general and MIMO CR in particular  is  that the SU must be able to mitigate the interference to the PU  blindly without any cooperation. Yi \citep{YiNullspace}
proposed a solution  in  the   case where the SU learns the channel matrix based on  channel reciprocity between the PU where the SU listens to the PU transmitted signal and estimates
 $\Hmat_{12}$'s
null space  from the signal's  second order statistics. This work was enhanced by Chen et. al. \citep{ChenInterference2011}.
 Both works  requires channel reciprocity and therefore are restricted to a PU that uses Time Division Duplexing (TDD).
Once the SU obtains the null space of \(\Hmat_{12}\), it does not interfere with the PU as long as his signal occupies that null space.

 Other than in the channel reciprocity case, obtaining the value of $\Hmat_{1j}$ by the SUs (i.e. the  interference channel to the PU) requires cooperation from the PU in the   estimation  phase, e.g. where the SU transmits  a training  sequence, from which the PU  estimates
 $\Hmat_{1j}$
 and feeds it back to the  SU.  Cooperation of this nature     increases system complexity overhead, since it  requires  a handshake between both systems  and in addition,   the PU needs to  be synchronized to   the SU's training sequence.
This is one of the major  technical obstacles that  prevents underlay CRs from being widespread.

\begin{figure}
\centering
\psfig{figure=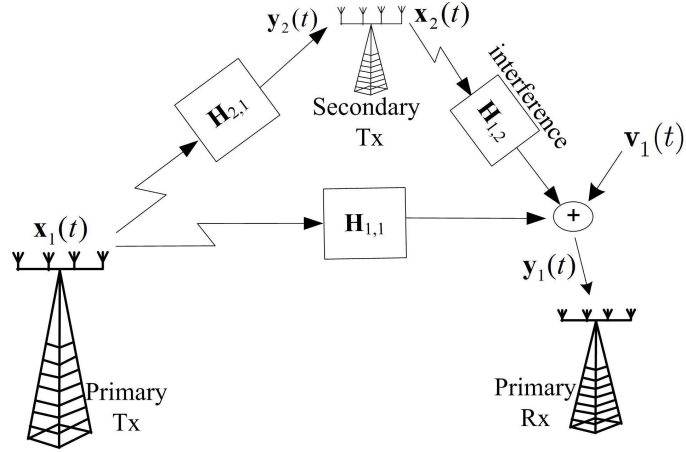 ,width=9cm}
\caption{  The  addressed cognitive radio scheme. The matrix   $\Hmat _{1,2}$ is unknown to the secondary transmitter and $\vvec_{1}$ (t) is a stationary noise   (which may include  stationary interference).   The  interference from the SU,
 ${\bf H}_{12}\xvec_2(t)$,
 is  treated  as noise, i.e. no interference cancellation is performed. The SU measures the energy variations  in
 $\yvec_{2}(t)$
 according to which it varies its transmission scheme until the interference to the PU
becomes sufficiently small such that it does not affect
$\yvec_2(t)$.}

\label{Figure:PassivePrimarysystem} \end{figure}

 The  objective of this work  is to design a simple procedure such  that MIMO underlay SU can  meet the interference constraint  to the   PU without explicitly estimating the matrix
$\Hmat_{1j}$
 and without burdening the PU with any handshaking, estimation or synchronization associated with the  SUs. In the proposed scheme (see \rm Fig.
 \ref{Figure:PassivePrimarysystem})  the PU is not cooperating at all with the  SU  and operates  as if it is the only system in the medium (as current PUs operate today).  The only condition required is that for some short time interval (that may be much shorter than the entire learning process), the PU  will be    using a power control scheme such that  its transmitted power is a monotonic function of the interference inflicted by the SU.  Under this condition, we propose a learning algorithm in which the   SU is   gradually reducing  the  interference to the PU  by iteratively modifying the spatial orientation of its   transmitted signal and measuring  the effect of this modification on  the PU's total transmit power.
The entire process is based  on energy measurements and on detecting energy variations.  Therefore, it  does not require  any handshake or  synchronization between the PU and the SU.

The paper is organized as follows: Section
\ref{Section:ProblemFormulation} provides the  system description and  some notation. Section \ref{Mathematical Descreaption of the Blind Jacobi Technique} presents a blind approach for realizing the Cyclic Jacobi technique  for calculating the  Eigenvalue Decomposition (EVD) of an unobserved  matrix; this will be the building block of the blind null space learning algorithm  presented in Section \ref{Section:BNSL Algorithm} and analyzed in  Section \ref{Complexity and Convergence}.  Simulations and  Conclusions are  presented  in  Sections \ref{Section:Simulation} and \ref{Section:coclusions},  respectively.

\section{ Problem Formulation}
\label{Section:ProblemFormulation}
\begin{figure}
\centering
{ \psfig{figure=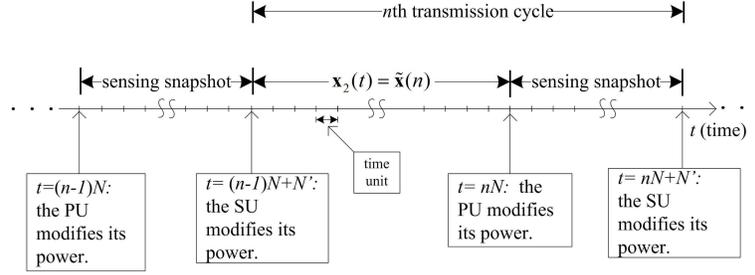,width=10cm}}
\caption{
The time indexing that is used in this paper. $t$  indexes  the basic time unit (pulse time) where \(N\) time units constitute a transmission cycle that is indexed by $n$.
Furthermore,  \(K\) transmission cycles constitute a learning phase (this is not illustrated in this figure).}
\label{Figure:index}
\end{figure}

Consider a flat fading MIMO interference channel  with  a single PU and a single  SU without interference cancellation, i.e. each system treats the other system's signal as noise. The PU's received signal is given by
\beq
\yvec_1(t) = \Hmat_{11} \xvec_1(t) + \Hmat_{12}\xvec_{2}(t) +\vvec_1(t),\; t\in {\mathbb N}
\eeq
  where  $\vvec_1(t)$ is a zero mean stationary noise. In this paper all vectors are column vectors. Let $\Amat$ be an $l\times m$ complex matrix, then, its null space is  defined as  $ {\cal N}(\Amat)=\{\yvec\in \mathbb C^{m}:\Amat \yvec=\0 \}$ where $\0=[0,...,0]^{\rm T}$.  For  simplicity $\Hmat_{12}$ will be denoted by  $\Hmat$
 and the matrix $\Hmat^*\Hmat$ will be denoted  by $\Gmat$. The time line $\mathbb N$ will be   divided into  $N$-length intervals  referred to as transmission cycles where, for each cycle, the SU's signal  will be constant (this is required only during the learning process),  i.e. \beq\label{SecondarySignal}
 \begin{array}{lll}\xvec_{2}((n-1)N+N') =\xvec_{2} ((n-1)N+1) \\~~~~~~~~~~~~=
\cdots=\xvec_2(Nn+N'-1)\triangleq\tilde
\xvec(n),\end{array}\eeq where the time interval  $nN<t\leq nN+N'-1$ ($N'<<N$) is the snapshot in which the SU measures  a function    \beq
\label{energyQ}
q(n)=\frac{1}{N'}\sum_{t=Nn}^{Nn+N'-1}\Vert \yvec_{2}(t)\Vert^{2}
\eeq
  where ${\bf y}_{2}$
is the  observed signal at the secondary transmitter  that  includes   the   primary  system's  transmitted  signal (see Figure
  \ref{Figure:PassivePrimarysystem}).   The SU's objective is to   learn ${\cal N}(\Hmat)$ from   $ \{\tilde\xvec(n),q (n)\}_{n\in {\mathbb N}}$.
 This  learning  process is carried out in learning stages where each stage consists  of $K$ transmission cycles. We will index each learning stage by $k$. The indexing method is depicted in Figure \ref{Figure:index}.
During the learning process, the SU varies the interference to the PU by transmitting a different interfering signal $\tilde\xvec(n)$.
    The  secondary transmitter measures  $\yvec_2(t)$ from which it extracts \(q(n)\) in order to  monitor the PU's transmitted power, i.e.
 $\frac 1 {N'}\sum_{t=nN}^{Nn+N'-1}\Vert \xvec_{1}(t)\Vert^{2}$.
 Each transmission cycle,
$N$,  corresponds to the PU's power control cycle, i.e.  to the  time interval between two consecutive power adaptations made by the PU.  In fact the actual $q(n)$ is not important as long as it satisfies  the following  condition.
\begin{assumption}
\label{Assumption:MonotonicAssumption}
Let $k\in {\mathbb N}$, then for every  $K(k-1)\leq n< kK$ the function $q(n)$ satisfies $q(n)=f_{k}(\Vert \Hmat\tilde\xvec(n)\Vert^{2})$ where $f_{k}:{\mathbb R_{+}}\rightarrow {\mathbb R_{+ }}$ is a sequence of strictly  monotonous continuous increasing  (decreasing) functions.
\end{assumption}
Without loss of generality, we assume that \(f_{k}\) is a sequence of monotone increasing functions.
The most important  trait (that will be  used in this paper) that follows  from Assumption
 \ref{Assumption:MonotonicAssumption}
is  that for every   $K(k-1)\leq n,m< kK$, $f_{k}(\Vert \Hmat \tilde\xvec(n) \Vert^{2})\geq f_{k}(\Vert \Hmat \tilde\xvec(m) \Vert^{2} )$ implies that $\Vert \Hmat\tilde \xvec(n)\Vert^{2}\geq \Vert \Hmat\tilde \xvec(m)\Vert^{2}$.
The problem is  illustrated in Figure \ref{Figure:BlindLearning}.
 From a system point of view, Assumption
 \ref{Assumption:MonotonicAssumption}
 means that between two  consecutive transmission  cycles  the   primary transmitter's  power can be   modified only due to   variations in the  interference level from the   SU that is at the learning process and not in a  steady state.

In the following we provide an example for conditions under which   Assumption \ref{Assumption:MonotonicAssumption} is satisfied.   To simplify the exposition, we replace the function $q$ in
\eqref{energyQ} with
\beq
\label{PracticalEnergyQ}
q(n)=\frac{1}{N'}\sum_{t=nN}^{Nn+N'-1} \E\{\Vert \yvec_{2}(t) \Vert^{2}\}
 \eeq
In doing so, we ignore the measurement noise at the secondary transmitter.  Such \(q(n)\)  satisfies  Assumption \ref{Assumption:MonotonicAssumption} if for example  \begin{enumerate}
\item The  PU's signal  has the following form:
 ${\bf x}_1(t)=\sqrt{p_1(n)}{\bf  P}_1 {\bf s(t)}$ where  ${\bf P}_1$ is a constant pre-coding matrix,${\bf }$ $\text{E}\left\{ \mathbf{s}(t)
\mathbf{s}{}^{*}(t) \right\}=\mathbf{I}$, and $p_{1}(n)$ is  the PU's power level at the $n^{\rm th}$ cycle.
This is satisfied for example if the PU  is using  beamforming  (in this case ${\bf s}(t)$ is a scalar and $\Pmat_1$ is a rank 1 matrix), or using a uniform power allocation (in this case ${\bf P}_1={\bf I}$) or if the PU  has  a    single antenna at the  transmitter.
 \item If in addition to condition 1 the PU has an SNR constraint at the receiver  or a constant  rate constraint.  Than  \(p_{1}(n)\) will be a monotonic increasing function of its total noise plus interference.
\item Assume  that condition 1 is satisfied and in addition the  PU is using OFDMA in which the current channel is just one of the bins and assume that the PU is using the  water filling rule. In that case, \(p_{1}(n)\) will be a decreasing function of the interference plus noise.
\end{enumerate}
Under such conditions  and assuming that \(\Hmat_{21}\) is unchanged between two transmission cycles, we have.
\beq
\begin{array}{lll}
  q_{p}(n)=\E\{\Vert\yvec_{2}(t)
  \Vert^{2}\}\\ \;\;\;\;\;\;=p_{1}(n)
  \E\{\Hmat_{21}
  \Pmat_1\svec(t)\svec(t)^{*}\Hmat_{21}
  \Pmat^{*}_{1}\}+\E\{\Vert\vvec_{2}(t)
  \Vert^{2}\}\\ \;\;\;\;\;\; =p_{1}(n)
  \Tr\{\Hmat_{21}
  \Pmat_1\Hmat_{21}\Pmat^{*}_{1}\}+
  \E\{\Vert\vvec_{2}(t)\Vert^{2}\}\\
  \;\;\;\;\;\;=a_1p_1(n)+a_2
\end{array}\eeq

The learning process is carried out as follows: At the first   cycle ($n=1$),  the SU transmits a low-power signal $\tilde \xvec(1)$,  such that the interference constraint
\eqref{InterferenceConstraint}
 is satisfied\footnote{
This can be obtained by transmitting a very low-power signal at the first cycle. If $q_{p}(n)$ is not affected, the power of $\tilde \xvec $ can be gradually increased until  $q(n)$ is affected.
}
 and measures the PU's transmit energy $q(1)$.  At the next cycle, the SU transmits   $\tilde\xvec_2(2)$   and  measures $q(2)$ and so on.  Section
 \ref{Section:BNSL Algorithm}
 describes algorithms for the SU to learn the null space of the interference channel matrix $\Hmat$ based on these measurements, i.e. to approximate
  ${\cal N}(\Hmat)$ from   $ \{\tilde\xvec (n),q_{} (n)\}_{n=1}^{T}$
where the accuracy can be arbitrary small for sufficiently large  $T$.  The algorithm is  not limited to networks with  a single SU; it is   also valid for networks with multiple SUs as long as only one  system  modifies  its power allocation during its learning process. This fact enables a new SU to join the channel in which       multiple SUs coexist with the PU  in a steady-state (i.e. a case where each SU meets its interference constraint given in
  \eqref{InterferenceConstraint}).

\section{ Blind Jacobi Diagonalization}
\label{Mathematical Descreaption of the Blind Jacobi Technique}
In this section we present a blind approach for realizing the Jacobi technique  for calculating the  EVD of an unobserved  Hermitian matrix $\Gmat$ assuming  that only
 $S(\Gmat,\xvec)=\xvec^*\Gmat\xvec$  is observed. This algorithm will be the building block of the blind null space learning algorithm that is  presented in Section
 \ref{Section:BNSL Algorithm}.
\subsection{The Jacobi Technique}
The Jacobi technique obtains the  EVD of the Hermitian  matrix $\Gmat$ via a series of 2-dimensional rotation that eliminates two  off-diagonal elements in each phase (indexed by $k$). It begins with setting     $\Amat_{0}
=\Gmat$ and then performing  the following  rotation operations
\beq\label{sigma_ {k}}\Amat_{k+1}= \Vmat_{k} \Amat_{k} \Vmat_{k}^*, k=1,2...\eeq  where $\Vmat_{k}=\Rmat _{l,m} (\theta,
 \phi)$  is  an $n_{t} \times n_{t}$ unitary rotation  matrix whose $p,q$ entry is given by:
\beq \label{Rlm(theta,phi)}[\Rmat_{l,m}
(\theta,\phivec) ]_{p,q}=
\left\{\begin{array}{lll}
  \cos(\theta ) & \text{if } & p =q\in \{m,l\} \\
 e^{-i \phi } \sin(\theta) & \text{if} & p=m \neq q=l \\
 -e^{i\phi}\sin(\theta ) & \text{if} & p=l \neq q=m \\
 1 & \text{if} & p=q\notin\{m,l\} \\
 0 &   & \text{otherwise}
 \end{array}\right.
\eeq
For each $k$, the vales of $\theta,\phi$ are chosen such that
$[\Amat_{k+1}]_ {l,m}=0$,
 in words, $\theta$ and $\phi $ are chosen to  zero $\Amat_{k}$'s  $l,m$ and  $m,l$ off diagonal entries. The  values of $l,m$ are chosen in step $k$ according to a function
 $J:{\mathbb N}\longrightarrow \{1,...,n _{t}\} \times \{1,...,n_{t}\} $    i.e $J_k=(l_{k}, m_{k})$. It is the choice of $J_{k}$
  that differs between different Jacobi techniques. For example, in the classic Jacobi technique, the off diagonal elements are chosen according to
  \beq(l_{k},m_{k})= \mathop {\arg \max }\limits_{\tiny \{(l,m):l>m\}} \left(\vert[ \Amat_{k }]_{l,m} \vert
\right)\eeq
 which corresponds  the  maximal off-diagonal entry\footnote{ Recall that $\Amat$ is Hermitian therefore it is sufficient to restrict $m>l$.}.
In the   cyclic Jacobi method the rotation rule is defined as follows:
\begin{definition}  $J_{k}=(l_{k
},m_{k})$  is a function such that    $1<l_{k}<n_t-1$ and $l_{
 k}<m_{k}\leq n_{t}$ where  each pair $(l,m)$ is chosen once in each cycle.
Unless otherwise stated it is assumed that \(l_{k}\leq l_{q}\) if \(k\leq q\) and that \(m_{k}\leq m_{q}\) if \(k\leq q\) and \(l_{k}=l_{q}\). \end{definition}
 For example if  $n_{t}=3$ then  $J_1 =(1,2) ,\;J_2=(1,3),J_3 =(2,3),\;
 J_4=(1,2)...$.


 \begin{figure}
\centering
{ \psfig{figure=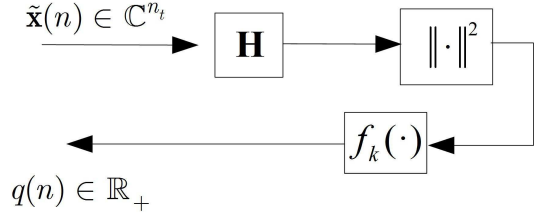,width=7cm}}
\caption{Block Diagram of the Blind Null Space Learning Problem. The SU's objective is to learn the null space of $\Hmat$ by inserting a series of $\{\tilde\xvec(n)\}_{n\in {\mathbb N}}$  and measuring \(q(n)\) as output.  The only information that can be extracted  is that $\Vert\Hmat
\tilde\xvec(n )\Vert^{2}\geq \Vert\Hmat \tilde\xvec(l )\Vert^{2}$ if $q(n)\geq q(l)$ for every $(k-1)K \leq l,n\leq kK$ where $k\in {\mathbb N}$.}
\label{Figure:BlindLearning}
\end{figure}
 \subsection{Blind Jacobi Step}
 In the proposed cognitive radio scenario, if the SU wishes to  perform the Jacobi step  it has to do so without observing $\Gmat$ and $\Amat_ {k}$ but observing only $f_{k}(\| \Hmat\xvec(n) \|^{2})=f_{k}
(\xvec^*(n)\Gmat\xvec(n))$  as depicted in Figure \ref{Figure:BlindLearning}. The following proposition shows how to do this:
\begin{theorem}\label{Theorem:blind step Theorem}
   Let $\Gmat$ be an $n_{t}\times n_{t}$ Hermitian matrix, let
 $\Rmat_{l,m}(\theta,\phi)$ be  defined in Eq. \eqref{Rlm(theta,phi)}, and let $S(\Gmat,\avec)=\avec^*\Gmat\avec$.   The  values of $\theta$ and $\phi$  that eliminate the $l,m$ entry of   $\Gmat$ i.e.\beq [\Rmat_{l,m} (\theta,\phi) \Gmat\Rmat_{l,m}^* (\theta,\phi)] _{l,m}=0 \eeq
 are given by
\beq\label{optimal Jacobi parameters}
(\hat\theta,\hat\phi)=\arg {\rm ext
}_{\theta,\phi} S\left(\Gmat ,\rvec_{l,m}
(\theta, \phi)\right)
\eeq
where \({\rm argext}\) denotes an extreme point and  $\rvec_{l,m}(\theta,\phi)$ denotes the \(l\)th column of  $\Rmat_{l,m} (\theta ,\phi)$.
 Furthermore, every local minimum point  of the function  $S\left(\Gmat ,\rvec_{l,m}(\theta, \phi)\right) $ is also an absolute minimum point., i.e. , let $\Gamma^*=\left\{(\theta^*,\phi^*) =\gammavec^* \in{\mathbb R}^{2}:\right.$ $\exists\; \epsilon>0$ $, S\left(\Gmat\right.$ $\left. \rvec_{l,m}(\gamma^*) \right)\leq S\left(\Gmat ,\rvec_{l,m}(\gamma)\right),$ $\left.\forall\;\Vert\gammavec- \gammavec^*  \Vert <\epsilon \right\}$, then $S(\Gmat,\rvec_{l,m}(\gammavec _{1}))=
S(\Gmat,\rvec_{l,m}(\gammavec_{2})),\; \forall\gammavec_{1},\gammavec_{2} \in\Gamma^*$.
 The same statement applies to local maxima.\end{theorem}
\IEEEproof See Appendix
 \ref{AppendixProofOfTheBlindStepTheorem}.

Theorem \ref{Theorem:blind step Theorem} asserts that the Jacobi step can be carried out by  a blind two-dimensional optimization in which every local minimum is a global minimum. This is a very important property since it is easy to identify if an  optimization algorithm has converged to a  local minimum.
Note that Theorem \ref{Theorem:blind step Theorem} applies also if $S(\Gmat ,\xvec)$ is not observed but  $f_{k}(S(
\Gmat,\xvec)$ is observed instead.


\subsection{ Reducing the Two-Dimensional Optimization into Two One-Dimensional  Optimizations}
\label{ReducingTheLineSearchComplexity}
Although the Jacobi step can be implemented blindly, it requires a two dimensional optimization over the parameters $\theta$ and $\phi $. This may be very difficult in practice since each  search point  is obtained via a transmission cycle. Fortunately, this two dimensional optimization  can be  carried out optimality by two one-dimensional optimizations as  shown in following theorem:
\begin{theorem}
 \label{Theorem:TweDimensionalSearchToOneDemenssionalSearch}
Let $S(\Gmat,\xvec)=\xvec^*\Gmat \xvec$ and let  $\rvec_{l,m} (\theta,\phi)$ be $\Rmat_{l,m}(\theta,\phi)$'s $l^{\rm th}$ column where $\Rmat_{l,m}(\theta,\phi)$ is  defined in  \eqref{Rlm(theta,phi)}. The optimal Jacobi parameters $\hat \theta$ and $\hat \phi$  in \eqref{optimal Jacobi parameters}
can be achieved by
\beqna
\label{ThetaOptimization}\hat\phi &=& \mathop {\arg\min  }\limits_{\tiny \phi  \in[-\pi,\pi ]} S\left(\Gmat , \rvec _{l}(\pi/3, \phi)\right)\\ \label{optimal phi}
\hat\theta&=&\left\{ \begin{array} {lcl} \tilde \theta & {\rm if}\;-\pi/4\leq\hat \theta\leq\pi/4 \\\tilde\theta-{\rm sign(\tilde \theta)}\pi/2 &{\rm otherwise}
\end{array} \right.\eeqna
where \({\rm sign}(\theta)=1\) if \(\theta>0\) and \(-1\) otherwise
and \beq
\label{tilde phi Optimization}
\tilde\theta=\arg\min_{\theta\in[-\pi/2,\pi/2]} S\left(\Gmat, \rvec_{l,m} (\theta,\hat\phi)\right)
\eeq
\end{theorem}
\IEEEproof See Appendix
\ref{AppendixProofOfTheBlindStepTheorem}.\\
{\bf Comment:} Note that the rotation  angle \(\theta_{k}\) is restricted to the interval \([-\pi/4,-\pi/4]\). In Section
 \ref{SectionConvergenceAnalysis} it is shown that this restriction  guarantees a globally linear convergence rate and ultimately a quadratic convergence rate.

In practice, the optimizations in \eqref{ThetaOptimization}, \eqref{tilde phi Optimization} will be carried out using  line searches. This is because, the function $S(\Gmat,\xvec)$ will   not be observed\footnote{In  \citep{NoamExploiting2011} it is shown that $S(\Gmat,\xvec)$ is known, the problem can be simplified drastically where \(\Gmat\) can be obtained precisely by a finite number of transmission cycles.} but only $f_{k}(S(\Gmat,\xvec))$  will be observed (see Figure
 \ref{Figure:BlindLearning}).
   The complexity of the  one dimensional  line search in \eqref{ThetaOptimization} can be  drastically reduced if one  is looking for a minimum (or maximum) and if   the objective function has a single minimum (local and global). Under these conditions  a  binary search can be invoked.
This is important since  it reduces the complexity of the line search drastically from \(O(1/\eta)\) to \(O(\log(1/\eta))\) where \(\eta \) is the line search accuracy.
 In the sequel, it will be  shown how to  solve both
 \eqref{ThetaOptimization} and  \eqref{tilde phi Optimization} using a binary search. We begin with the optimization in \eqref{tilde phi Optimization} which can be written
as
\beq\label{optimizeQA}
\tilde\phi_{k}=\arg\min_{\phi\in[- \pi,\pi]} f_{k}\left(S\left(\Gmat, \rvec_{l,m} (\pi/3,\phi)\right)\right)
\eeq
Note that \beq
\label{TrigonometricS}
\begin{array}{lll}
S(\Gmat,\rvec_{l,m}(\theta,\phi))=\cos ^2(\theta ) \left|g_{l,l}\right| +\sin ^2(\theta ) \left|g_{m,m}\right|
\\~~~~~~~~~~~~~~~~~~~~~
~~-\left|g_{l,m}\right| \sin (2 \theta ) \cos ( \phi+\angle g_{l,m} )
\end{array}\eeq
  It follows that during each learning phase, that is for every   $(k-1)K\leq n\leq kK$, the function  $f_{k}
\left(S(\Gmat ,\rvec_{l,m}
(0.5,\phi))\right)$ is equivalent to
\beq\label{define w}
w_{k}(\phi)=f_{k}(A+B\cos(\phi+\angle g_{l,m}))\eeq
where $f_{k} $ is a monotone function and  $A\geq \vert B\vert $.
 \begin{proposition} \label{antisymetric} Consider the function $w$ in \eqref{define w} then
 \begin{enumerate}[a. ]
  \item $w(\phi)$ is monotonic  on $[\pi/2,\pi]$ and non-monotonic on $(0,\pi/2)$  if and only if
 \beq\label{monotone condition}\vert w(0)-w(\pi/2)\vert\leq\vert w(\pi/2)
-w(\pi)\vert\eeq
 \item  Let $\check\phi\in(- \pi,\pi)$ be a  minimum\footnote{Since $w$ is assumed monotone and continuous and its argument  is a $2\pi$  periodical sinusoid such point always exist if $w(0)\neq w(\pi)$.  } point of $w(\phi)$ then:   \begin{enumerate}[i.] \item    Assume \eqref{monotone condition} is true, then, $\check \phi \in[0,\pi/2]$ if $w(\pi/2)\geq w(\pi)$ and $\check \phi\in[- \pi, -\pi/2]$ if $w(\pi/2)\geq w(\pi)$. \item Assume  \eqref{monotone condition} is false, then,  $\check \phi\in[-\pi/2,0]$ if $w(\pi/2)>w(0)$ and $\check \phi\in[\pi/2,\pi]$ if $w(\pi/2)<w(0)$.         \end{enumerate}  \end{enumerate} \end{proposition}

\IEEEproof
This follows immediately from the fact that  $A\cos( \phi-\angle g_{l,m})+B$ is affine symmetric\footnote{A function $f:{\mathbb R} \longrightarrow {\mathbb R}$ is affine symmetric if there exist some  $ a\in{ \mathbb R}$ such that $f(x-a) =f(a-x)$.}    in the intervals $[\angle g_{l,m}-\pi/2,\angle g_{l.m}+\pi /2]$ and $[\angle g_{l,m}+ \pi/2,\angle g_{l.m}+3 \pi/2]$ with unique extreme point.  \hfill $\Box$

Proposition \ref{antisymetric} can determine, via $3$ points, a   $\pi
/2$-length closed-interval in which $\check \phi=\angle g_{l,m}$ is the unique local (and therefore global) minimum. It is now possible to invoke a binary search to approximate $\angle g_{l,m}$ by $\hat \phi$ where for any  accuracy, say  $\eta>0$, it takes an additional  $-\log_{2}(2 \eta/\pi)$ points (transmission cycles)  to ensure $\check \phi\in[\angle g_{l,m}-\eta ,\angle g_{l,m} +\eta]$. In exactly the same manner it can be shown that the optimal value of $\hat \theta $ in
 \eqref{ThetaOptimization} can be approximated $\eta$ closely  using  $3-\log_{2}(\eta/\pi )$ transmission cycles. The algorithm is summarized in Table \ref{Line Search Table}.
\begin{table}
\caption{Line Search}
\label{Line Search Table}
function: $z={\rm \bf LineSearch} (w,z_{\min},z_{\max},\eta)$
\begin{enumerate}
\item
Initialize: $z=z_{\max}/2$; $a,b,c=0$; $L=z_{\max}$.
\item If $ \left|w\left(z_{\max }/2
\right)-w(0)\right|>\left|w\left(z_{\max }/2\right)-w\left(z_{\max }\right)\right|$, set $a=1.$
\item If $w(0)<w\left(\left.z_{\max }\right/2\right)$, set $b=1$.
\item If $w\left(z_{\max }\right)
>w\left(\left.z_{\max }\right/2\right)$ set $c=1$.
\item
$z_{\max }=z_{\max } a (1-b)+z_{\max } (1-a) c;$
\item $z_{\min}=z_{max}-L$.
\item While $\left(\vert z_{\max}
-z_{\min}\vert\geq \eta\right)$
\item $z=(z_{\max}+z_{\min})/2$
\item If $(w(z_{\rm min})
\leq w(z_{\max}))$, set $z_{\max}=z$, otherwise, set $z_{\min}=z$.
\item  end while
\end{enumerate}
end ${\bf LineSearch}$\end{table}

\section{ Blind Null Space learning via the cyclic Jacobi Technique }
\label{Section:BNSL Algorithm}
In this section we present a Blind Null Space Learning (BNSL) algorithm  for solving the blind null space learning  problem based on  Theorems \ref{Theorem:blind step Theorem}-
\ref{Theorem:TweDimensionalSearchToOneDemenssionalSearch}  and on the line search algorithm in Table \ref{Line Search Table}. The SU  is using a pre-coding  matrix that is being  updated at each learning stage, let   $\Wmat _{k}$ be that  pre-coding matrix   at stage $k$.
\subsection{The  Blind Null Space Learning (BNSL) Algorithm}
Let $\Umat \bSigma \Vmat^*$ be  $\Hmat$'s   Singular Value Decomposition, where $\Vmat$  and $\Umat$ are  $n_{t}\times n_{t}$ and $n_{r} \times n_{r}$ unitary matrices respectively and assume that \(n_{t}>n_{r}\). The matrix   $\bSigma$ is an $n_{r}\times n_{t}$ diagonal matrix with real nonnegative  diagonal entries $\sigma_{1} ,...,
 \sigma_ {d}$  that are  arranged as  $\sigma_{1}\geq\ \sigma_ {2},\geq \cdots\geq\sigma_ {d}>0$. We assume with no loss of generality that \(n_{r}=
d(={\rm Rank}(\Hmat))\). Note that \beq
{\mathcal N}(\Hmat)={\rm span} \left (\vvec_{n_{r}+1} ,...,\vvec_ {n_{t}}
\right)
\eeq
where $\vvec_{i}$ denotes $\Vmat$'s $i$th column. From the SU point of view, it is sufficient to learn ${\cal N}(\Gmat)$ (which is equal to ${\cal N}(\Hmat)$ because $\Gmat= \Vmat \bLambda \Vmat^*$ where $\bLambda=
\bSigma ^{\rm T}\bSigma)$.  It is  possible to diagonalize $\Gmat$ via the blind Jacobi algorithm given in Table \ref{Blind Jacobi table}. The secondary user's initial   (i.e. at \(k=0\)) pre-coding matrix  is \(\Wmat_{0}=\Imat\). Each Jacobi   step  is  equivalent to a learning stage   (indexed by $k$, see Figure \ref{Figure:index}) and is composed of $K$ transmission
cycles where the secondary user obtains  \((\hat\theta_ {k}, \hat\phi_{k})\) at its end,  and updates  its pre-coding matrix
\beq\label{Defind:W_k}\Wmat_{k}=\Wmat_{k-1} \Rmat_{l_{k},m_k} (\hat\theta_k,\hat \phi_k)\eeq Obtaining \((\hat\theta_ {k}, \hat\phi_{k})\) at each learning  stage
 requires two binary searches where
the value of $f_{k} (S(\Gmat ,$ $\Wmat_{k} $$\rvec_{l_{k}, m_{k}} (\theta(n),\phi(n))))$ for each search point ($\theta_{n}, \phi_{n}$) is obtained by a transmission
cycle in which the SU is transmitting       \beq \label{define x}
      \begin{array}{lll} \xvec_{2} (t)=\tilde \xvec(n)= \Wmat_{k-1} \rvec_{ l_{k},m_{k}} (\theta_n,\phi_n)\in{ \mathbb C}^{n_{t}} \\~~~~~~~~~~~ ~~~~~~~ ~~~~~~ ~~~~~~~~,\; \forall (n-1)N\leq t\leq nN\end{array}\eeq i.e. for each $k$, the SU  performs two one-dimensional binary searches to obtain  $(\hat\theta_{k} ,\hat \phi_{k})$ (the minimum of  $f_{k }(S(\Gmat,\Wmat_{k-1} \rvec_{l_k,m_{ k}} (\theta,\phi)))$) from the set $\{\rvec_{l_{k},m_{k}}(\theta_{n}, \phi_{n}) ,$ $f_{k}(S (\Gmat, $  $\Wmat_{k-1} \rvec_{l_k,m_{k}} (\theta_{ n}$ $,\phi_{n})))\}_{n=(k-1)K}^{kK-1}$. As the algorithm proceeds,    \(\hat\theta_{k}\) approaches zero (this will be shown on Section \ref{Complexity and Convergence}), a fact that can be used for a stopping criteria.

Assume that the BNSL algorithm is performed until  \(k=k_{s}\). Then the SU  pre-coding matrix \(\Tmat_{k_{s}}\) is given by
\beq
\label{DfinePks}
\Tmat_{k_{s}}=[\wvec^{k_{s}}_{i_{1}},...,
\wvec^{k_{s}}_{i_{n_{t}-n_{r}}}]
\eeq
 where \(\wvec^{k}_{i}\) is \(\Wmat_{k}\)'s \(i\)th column and  \(i_{1},i_{2},...,i_{n_t}\) is an indexing such that
\beq
(\wvec^{k_{s}}_{i_{1}})^{*}\Gmat \wvec^{s}_{i_{1}} \leq (\wvec^{k_{s}}_{i_{2}})^{*}
\Gmat\wvec^{s}_{i_{2}} \leq \cdots \leq (\wvec ^{ k_{s}}_{i_{n_{t}}})^{*} \Gmat\wvec^{s}_{i_{n_{t}}}
\eeq
Thus, the interference power that the SU inflict  to the PU is bounded \beq \Vert \Hmat_{12} \xvec_{12}\Vert^{2}\leq p_{2}\Vert  \Hmat_{12} \Tmat_{k_{s}}\Vert^{2} \eeq where $p_2$  is the SU's transmit power.
In Section \ref{Complexity and Convergence} it will be shown that \(\Vert \Hmat_{12} \Tmat_{k}\Vert^{2}\) converges quadratically to zero as is bounded by for sufficiently small \(\eta\) and ultimately bounded by \(O(\eta^{2})\).

\begin{table}
\caption{The Blind Null Space Learning (BNSL) Algorithm}
\label{Blind Jacobi table} Given the sequence $\{h_{k}\}_{k\in {\mathbb N}}$ defined in \eqref{Definh_k}\\  function: $\Wmat={\rm \bf BNSL}(\{h_{k }\}_{k\in {\mathbb N}},n_t)$
\begin{enumerate}
\item
Initialization: $k=1$,  $\Wmat_{0}=\Imat_{ n_{t}}$, $\Delta_{j}=2\xi, \forall j\leq 0$.
\item while $\left(\max_{j\in \{k-n (n-1)/2,...,k\}} \Delta_{j}\geq \xi \right)$.
\item $w(\phi)=h_{k}\left( \Wmat_{k-1} \rvec_{l_{k},m_{k}} (\pi/3,\phi) \right)$.
\item \label{line:BNSL4}set $\hat \phi_{k}={\rm \bf LineSearch}(w,0 ,\pi,\eta)$
\item \label{line:BNSL5}$w(\theta)=h_{k} \left(\Wmat_{k-1}\rvec_{l_{k},m_{k}} (\theta, \hat \phi_{k}) \right) $.
\item \label{line:BNSL6}set $\tilde \theta_{k}={\rm \bf LineSearch}(w,0 ,\pi/2, \eta)$
 \item set $\hat \theta_{k}$ according to \eqref{optimal phi}.
\item  set $\Delta_{k}= \vert \hat\theta_{k}\vert $
\item set $\Wmat_{k}= \Wmat_{k-1} \Rmat_{l_{k},m_k} (\hat\theta_k,\hat \phi_k),\; \Wmat=\Wmat_{k}$
\item $k=k+1$.
\item end while
\end{enumerate}
\end{table}

It is important to note   the eigenvalues of \(\Gmat \) cannot be obtained by the BNSL algorithm.  This  leaves the SU with the problem of how to determine which of the columns of \(\Wmat_{k_s}\) corresponds to  \(\Hmat_{12}\)'s null space, if it doesn't know its's rank in advance. The problem however can be solved  due to the fact that \(\vvec_{i}\in{\cal N}(\Gmat)\) iff \(S(\Gmat,\vvec_{i})=S(\Gmat,2\vvec_{i})\). Denote  $h_{k}:{\mathbb C}^{n_{t}}
\rightarrow {\mathbb R}_{+}$ as  \begin{equation}
\label{Definh_k} h_{k}(\xvec)=f_{k}(S(
\Gmat,\xvec )),\end{equation} thus, by setting \(\tilde\xvec(n)=\vvec_{i}\) and \(\tilde\xvec(n+1)=2\vvec_{i}\) it follows that \(\vvec_{i}\in{\cal N}(\Gmat)\) iff \(h_{k}(\tilde\xvec(n))=
h_{k}(\tilde\xvec(n+1))\). The same  approximately applies for $\Wmat_{k_s}$ when $k_s$ is sufficiently large.

\subsection{The  Reduced Complexity  Blind Null Space Learning (RC-BNSL) Algorithm}
\label{SectionConvergenceAnalysis}
In the BNSL algorithm, each  step  requires  two line searches where each search  point is obtained by a transmission cycle. These transmission cycles  are the dominant latency factor     in the learning process since  the rest of the calculations are performed off line at  the secondary device processing unit. Roughly speaking, the complexity of the cyclic Jacobi technique grows   like $n_{t}^{2}$ which is the dimension of the matrix  $\Gmat$ (the convergence and complexity will be  discussed in Section \ref{Complexity and Convergence}). In this Section we present an algorithm that converts the problem of blind null space learning of  the $n_{t}\times n_{t}$ matrix  $\Gmat$ into an equivalent problem of  blind null Space learning  of $n_{t} -n_{r}$ matrices  where each  is an  $n_{r}\times n_{r}$ matrix.   This is possible due to the  fact that $\Gmat=\Hmat ^*\Hmat$  is a  $n_{r}$-rank matrix and therefore has a  $n_{t}-n_{r}$ dimensional null space\footnote{If the matrix $\Gmat$ was known, the fact that ${\rm rank}(\Gmat) =n_{r}$ could have been utilized by QR decomposition (which cannot be done blindly) prior to the diagonalization. }. The resulting complexity grows like $(n_{t}-n_ {r})n_{r}^{2}$, therefore, the algorithm is efficient  if  $n_{t}$ is sufficiently larger than $n_{r}$ (which is a  practical case  since SU systems will be more sophisticated than the PU).

The idea behind the RC-BNSL algorithm is described in the  following observation:
\begin{observation}
Let $\Hmat\in\comp^{n_r\times n_{t}}$ be an \(n_{r}\)-rank  (\(n_{t}>n_{r}\)) matrix  and let  \(\Umat=[\mathbf{u}_{1 },...,\mathbf{u}_{n_{r}+1}]\in\comp^{n_{ t}\times (n_{r}+1)}\) be an orthonormal matrix (that is, a matrix whose columns are an orthonormal set i.e. \(\Umat^{*} \Umat=\Imat\)) and let \(\tilde \Hmat= \Hmat\Umat\). If \(\tilde \uvec\in {\cal N}(\tilde \Hmat)\) then \(\uvec= \Umat\tilde \uvec\in{\cal N}(\Hmat)\). \end{observation} \IEEEproof
This is due to \(\Hmat\uvec=\Hmat\Umat \tilde \uvec=\0\) since \(\tilde \uvec\in{\cal N}(\Hmat\Umat)\).

 The RC-BNSL algorithm is carried out as follows: The secondary user begins with an initial pre-coding matrix \(\Umat^{(0)}\in {\mathbb C}^{n_{r}\times (n_{t}-n_{r})}$ which is  composed of the last $n_{t}-n_{r}$ columns of some unitary  matrix $\Wmat\in \comp^{n_{t}\times n_{t}}$.  Let  $ \Hmat_{\rm eq}^{(1)}= \Hmat\Umat^{ (0)}\in\comp^{n_{r}\times( n_{r}+1)}$, then there exists at least  one degree of freedom in this channel.
 The SU can apply the BNSL algorithm on \( \Gmat^{(1)}=\Hmat^{(1)^{*}}_{\rm eq}\Hmat^{(1)}_{\rm eq}\) and obtains a pre-coding matrix \(\Umat_{k}^{(1)}\)  such that  $\tilde\Umat^{( 1)}=\lim_{k \rightarrow\infty}\tilde \Umat_{k}^{(1)}$ (in Section \ref{Complexity and Convergence} it is shown that the limit exists) and that \beqna
\bLambda^{(1)}
=\tilde\Umat^{(1)^{*}}\Gmat^{(1)} \tilde\Umat^{(1)}
=\begin{bmatrix}0 & 0 & \cdots & 0 \\
0 & \lambda^{(1)}_{n_{r}} & \ddots & \vdots \\
\vdots & \ddots & \ddots & 0 \\
0 & \cdots & 0 & \lambda^{(1)}_{1} \\
\end{bmatrix}
\eeqna
 Now that $\Gmat^{(1)}$ is diagonalized  the first degree of freedom is given  by \(\vvec^{(1)}= \Umat^{(0)}\tilde \uvec_{1}^{ (1)}\)  where \(\tilde \uvec_{1}^{(1)}\) is the first column of $\tilde\Umat^{(1)}$, (that lies in the null space of $\Gmat^{(1)}$). The SU then can gain an additional degree of freedom  by applying the BNSL algorithm on the following $(n_{1}+1)$ equivalent channel
\beq\Hmat_{\rm eq}^{(2)}=\Hmat\Umat^{(1)}
\eeq where $\Umat^{(1)}\in \comp^{n_{t} \times n_{r}+1}$ is obtained by concatenating the \(n_{r}-1\) column of the initial unitary matrix \(\Wmat\) with the last \(n_{r}\) column of \(\tilde \Umat^{(1)}\) multiplied by \(\Umat^{ (0)}\), i.e. let  \(\hat\Umat^{(1)}=[ \tilde \uvec_{2}^{(1)},...,\tilde \uvec^{(1)}_{n_{r}+1}]\) then
 $\Umat^{(1)}=[\wvec_{n_{r}-1},\Umat^{ (0)}\check\Umat^{(1)}]$.
This equivalent channel is then diagonalized using the BNSL algorithm to obtain $\tilde\Umat ^{(2)}$. We now have two degrees of freedom given by $\vvec^{(2)}= \Umat^{(1)}\tilde\uvec_{ 1}^{(2)}$and $\vvec^{(1)}$. This process is repeated until all \(\Wmat\) column are used.  The RC-BNSL algorithm is summarized in Table \ref{table 2}.
\begin{table}
\caption{The Reduced Complexity Blind Null Space Learning (RC-BNSL) Algorithm}
\label{table 2}
Let \(\Wmat\in\comp^{n_{t}\times n_{t}} \) be a unitary matrix. \\function: $[\vvec_{1} \cdots\vvec _{n_{t} -n_{r}}]={ \rm \bf RC\underline~ BNSL}(\{h_{k}\}_{k\in\nat },n_{r},n_{t})$
\begin{description}
\item Initialize \(\Umat^{(0)}=[\wvec_{n_t-n_r },...,\wvec_{n_t}]\).
\item for $m=1,\dots, n_{t}-n_{r} $
\begin{description}
\item $a(\xvec)=h_{k}(\Umat^{(m-1)}\xvec)$
\item $\tilde\Umat^{(m)}={\rm \bf BNSL}(a,n_{r}+1)$
\item $\vvec_{m}=\Umat^{(m-1)}\tilde \uvec_{1}^{(m)}$
\item $\check\Umat_{m}=[\tilde \uvec_{ 2}^{(m)},...,\tilde\uvec^{(m)}_{n_{r} +1}]$
\item $\Umat^{(m)}=[\wvec^{(m)}_{n_{t}- n_{r}-m}, \Umat^{(m-1)}\check \Umat^{ (m)}]$
\end{description}
\item end for
\end{description}
\end{table}

\section{Convergence }
\label{Complexity and Convergence}
In this section we study the convergence and complexity properties  of the BNSL algorithm. Recall that this  algorithm is  in fact a blind implementation of the cyclic Jacobi technique whose convergence properties  have been extensively studied over the last 50 years. However, while the  convergence properties  of the Cyclic Jacobi technique  directly apply to its   Blind implementation  due to Theorem \ref{Theorem:blind step Theorem}, they cannot be applied directly to the BNSL  algorithm. This is due to fact that in the latter algorithm      the rotation angles $\theta_{k}, \phi_{k}$ are approximated  via a line search of finite accuracy \(\eta \) for every $k$ (see Table \ref{Blind Jacobi table}, lines \ref{line:BNSL4}-\ref{line:BNSL6}) while in the previous, $\theta_{k},\phi_{k}$,  are equivalent (according to Theorem
\ref{Theorem:TweDimensionalSearchToOneDemenssionalSearch}) to the rotation angles of the Cyclic Jacobi technique.
  Moreover, we would like to make this line search accuracy as small as possible (that is, to make \(\eta \) as large possible) in order to reduce the number of transmission cycles. It is therefore very important to understand how  \(\eta\) affects the performance of the BNSL algorithm. In this section we will extend the classic convergence  results  of the Cyclic Jacobi technique to the   BNSL algorithm and show what is the required accuracy in the line search that assures convergence and a minimal level of interference inflicted by the SU to the PU.

 \subsection{Overview of the previous work on the convergence of the Jacobi technique}
\label{Literature Survay}The convergence of the Jacobi technique has been studied extensively over  the last sixty years.  
The first convergence proof of the Cyclic Jacobi technique for complex Hermitian matrices was given by Foster and Henrici \citep{forsythe1960cyclic}, which proved that if in each step $k$: 1)  The off diagonal entry satisfies $\vert a^{k}_{l_ {k},m_{k}}\vert<c_{k} \vert a^{k+1}_ {l_{k},m_{k}}\vert$, where $ a_{l,m}^{k}$ is \(\Amat_{k}\)'s \(l,m\)  entry  (defined in \eqref{sigma_ {k}}) and  $0\leq c_{k}\leq b<1$ where $b$ is independent of $k$.  2) The rotation angle (in this paper it is $\theta_{k}$ defined in
\eqref{ThetaOptimization})  lies in some close  interval $A\in(- \pi/2,\pi/ 2)$, then the cyclic technique converges to $\Gmat$'s EVD.
These conditions are satisfied by the BNSL algorithm for any line search accuracy for which \(\theta_{k}\neq 0\). Thus, to guarantee convergence of the BNSL algorithm to  \(\Gmat\)'s EVD, one should follow the following rule: If \(\theta_{k}=0\), phase \(k\) should be repeated with  \(\eta\) being  decreased  so that \(\theta_{k}\neq 0\).  If any further decrease in \(\eta\) does not change   \(\theta_{k}\neq 0\), then \(a^{k}_{l_{k},m_{k}}\)   is already zero. This however does not indicate  of the convergence rate and does not take into considerations that    \(\eta\) cannot be infinitely decreased.

Henrici, and Zimmermann \citep{ henrici1968estimate} proved that the cyclic Jacobi algorithm for real symmetric matrices has a global linear convergence rate\footnote{A sequence \(a_{n}\) is said to have a linear convergence  rate of \(0<\beta<1\) if \(\vert a_{n+1}\vert<\beta \vert a_{n}\vert \).} that depends on the matrix size $n_t$  if the rotation angle $\theta_{k}\in[-\pi/4,\pi/4]$ for every $k$.   Fernando \citep{FernandoNumerical1989} extended this result to complex Hermitian  matrices.
  A very   important result is the ultimate quadratic convergence\footnote{A sequence is said to have a quadratic convergence  rate if there exist some \(\beta>0\) such that \(\vert a_{n+1}\vert<\beta \vert a_{n} \vert^{2}\).} rate of the Cyclic Jacobi technique that was shown by Henrici \citep{HenficSpeed1958}   for complex Hermitian matrices with well separated eigenvalues and later enhanced by  Wilkinson \citep{WilkinsonNotes1962}.  The most recent and comprehensive result of the quadratic  convergence of Cyclic Jacobi technique that includes multiple and clusters of  eigenvalues is due to Hari \citep{VjeranSharp1991}. Once the Cyclic Jacobi algorithm  reaches its quadratic convergence rate it takes a very small number of cycles to reach any desirable accuracy, however, there is no rigorous bound on the number of the required cycles to reach that rate. Brent and Luk \citep{BrentSolution1985} have argued heuristically that this number is \(O(\log(n_{t}))\) cycles for  \(n_t\times n_{t}\)  matrices. This seems to be the case in practice \citep[page 429]{golubmatrix} where such a rapid decline is obtained   after three to four cycles \citep[page 197]{parlett1998symmetric}. For further reading the reader is referred  to \citep{Golub200035,golubmatrix,parlett1998symmetric}.

\subsection{Global Linear Convergence rate  }
  The global linear convergence  of the Cyclic Jacobi technique   was derived  by Fernando \citep[]{FernandoNumerical1989}:
\begin{theorem}[\citep{FernandoNumerical1989},Theorem 4]\label{FernandoTheorem}
Let $\Gmat$ be a finite dimensional $n\times n$ complex Hermitian matrix and $P_{k}$ denote the norm of the off diagonal upper triangular (or lower triangular) part of $\Amat_{k}$ (as defined in \eqref{sigma_ {k}}) and let $m=n(n-1)/2$ (which is a cycle length).    Then
\beq
\label{Linear convergence rate}
P_{k+m}\leq\rho_{}P_{k}, \rho= \left( 1-2^{-(n_{t}-1)(n_{t}-2)/2}  \right)^{1/2}
\eeq
\end{theorem}
if
 \beq \max_{v\leq k+m} \{\vert\theta_{v} \vert\}\leq \frac \pi 4\eeq

Although Theorem \ref{FernandoTheorem} is the tightest closed form   bound on the global convergence rate of the cyclic Jacobi technique,  the practical convergence rate is much faster as  discussed in Section \ref{Literature Survay}.

The condition of  Theorem \ref{FernandoTheorem}  is satisfied in  the BNSL algorithm  as we show in Theorem
\ref{Theorem:TweDimensionalSearchToOneDemenssionalSearch}.  However, as described in Table \ref{Blind Jacobi table}, the angles $\theta_{k}$ and $\phi_{k}$ are approximated via a line search. Thus  the off diagonal elements are not completely annihilated, i.e.   $[\Amat_{k}]_{l_{k},m_{k}}\approx0$ instead of $[\Amat_{k}]_{l_{k},m_{k}}=0$.  In the following Theorem we show what is the  required line search accuracy that guarantees the  convergence rate in \eqref{Linear convergence rate}.
\begin{theorem}\label{Proposition Linear Convergence} Let $\Gmat$ be a finite dimensional $n_{t}\times n_{t}$ complex Hermitian matrix and $P_{k}$ denote the norm of the off diagonal upper triangular (or lower triangular) part of $\Amat_{k}=\Wmat^{*}_{k-1}\Gmat\Wmat_{k-1}$ where \(\Wmat_{k}\) is defined in \eqref{Defind:W_k} (see also Table \ref{Blind Jacobi table})    and let $m=n_{t}(n_{t}-1)/2.$
Let $\eta$ be the accuracy of the line search,  then the BNSL algorithm has a globally linear convergence rate that is given
 \begin{equation}
 \label{EtaInequilityOfTheTheorem} \begin{array}{lll}P^{2}_{k+m} \leq P^{2}_{k}\left( 1-2^{-(n_{t}-2) (n_{t}-1) /2 }\right)+(n_{t}^{2}-n_{t})(7+2\sqrt 2 )\eta^{2}\Vert\Gmat\Vert^{2}
\end{array} \end{equation}
  \end{theorem}
\IEEEproof See Appendix \ref{Appendix3}.

To demonstrate this result
we substitute  \(\eta=a {P_{qm}} /\Vert \Gmat\Vert\) and  obtain
\beq
P_{k+m}\leq P_{k} \left(1-2^{\frac{ 1}{2} (2-n_{t}) (n_{t}-1)}+\left(7+2\sqrt{2}\right) a^2 \left(n_{t}^2-n_{t}\right)\right)^{ 1/2}
\eeq
It follows that for the BNSL to have a linear convergence rate, it is sufficient that the accuracy be
at least:\beq
\begin{tabular}{|c|c|c|c|c|c|} \hline
\(n\) & 3 & 4 & 5 & 6 & 8  \\\hline
\(\eta\times\frac{\Vert\Gmat\Vert}{ {P_{ k}}}\) & \(8\times 10^{ -2}\) & \(2\times10^{-2}\) & \(7\times10^{ -3}\) & \(1\times10^{ -3}\) & \(2\times10^{-5}\) \\\hline
\end{tabular} \eeq

Note that the linear convergence coefficient in Theorem \ref{Proposition Linear Convergence} may be very small for a large number of transmitting antennas \(n_{t}\). This may be very bad if this bound turns out to be tight, i.e. if \(\rho\) is very close to the true convergence coefficient.
In the sequel it will be shown that the actual convergence rate of the BNSL is much faster than the bound in Theorem \ref{Proposition Linear Convergence}.

\subsection{An Asymptotic  Quadratic  Convergence Rate}
So far it has  been  shown that for a right chose of \(\eta\), the BNSL algorithm converges and that for sufficiently small \(\eta\) it has at least a global linear convergence rate.   In the following theorem and corollary it is shown what is the desired \(\eta\) to guarantee an asymptotic  quadratic convergence rate and  what \(\eta\) guarantees  that the SU will meet the  maximal interference constraint of the PU.
\begin{theorem}
\label{TheoremQuadraticConvergence}
Let $\eta$ be the accuracy of the line search,   \(\{\lambda_{l}\}_{l= 1}^{n_{ t}}\) be
  \(\Gmat\)'s eigenvalues  and let
  \beq
  3\delta=\min_{\lambda_{l}\neq \lambda_{r }}\vert \lambda_{l}-\lambda_{r}\vert
  \eeq
    Let  $P_{k}$ be  the norm of the off diagonal upper triangular  part of $\Amat_{k}=\Wmat^{*}_{k-1}\Gmat\Wmat_{k-1}$ where \(\Wmat_{k}\) is defined in \eqref{Defind:W_k} (see also Table \ref{Blind Jacobi table})    and let $m=n_t(n_t-1)/2.$
Assume that the BNSL algorithm has reached a stage where \(P^{2}_{k}<\delta^{2}/8\), then
 \begin{equation}
 \label{EtaInequilityOfTheTheorem}
 \begin{array}{lll} P^{2}_{k+m}\leq O \left(\left( \frac {P_{k}^{2}}{\delta} \right)^{2}\right)+O \left( \frac {  \eta P_{k}^{3/2}}{\delta} \right) +O \left( \frac {  \eta ^{2}P_{k}^{1/2}}{\delta} \right) \\\;\;\;\;\;\;\;\;\;\;\;\;\;\;\;\;\;\;\;\;\;\;\;\;+ 2\left(2n_tn_{r}-n_r^{2}-n_r \right) \eta^{2}\Vert\Gmat\Vert^{2}
\end{array} \end{equation}.
  \end{theorem}
\IEEEproof  See Appendix
 \ref{ProofOfTheoremQaudraticConvergence}.

Theorem \ref{TheoremQuadraticConvergence} shows  that to guarantee quadratic convergence  rate the  accuracy should be    much smaller than \(P_{k}^{2}\), that is,   let \(k_{0}\) be an integer  such that \(P^{2}_{k_{0}}<\delta^{2}/8\) then  \beq P_{k_0+m}\leq O\left(\left(\frac{P_{k_0}}
{\sqrt\delta}\right)^{2}\right)\eeq
if   \(\eta<<P^{2}_{k_{0}}\). This implies that  once \(P_{k}\) becomes very small  such that    \(P^{2}_{k}<<\eta\),  one cannot guarantee that  \(P_{k+m}\) will be smaller than \(P^{2}_{k}\) but  only be smaller  than \(O(\eta)\). A fact that motivates derivation of a      bound on  the interference power that the SU inflict to the PU as a function of \(\eta\).
\begin{corollary}\label{corollary}
Let \(\Tmat_{k}\) be the SU's pre-coding matrix defined in \eqref{DfinePks}. Assume that the conditions of  Theorem
\ref{TheoremQuadraticConvergence} are satisfied, then
\beq\label{BoundOnInterference}
\Vert\Hmat_{12}\Tmat_{k+m}\Vert^{2}\leq O \left(\left( \frac {P_{k}^{2}}{\delta^{3/2}} \right)^{2}\right)+O \left( \frac {  \eta P_{k}^{3/2}}{\delta^{2}} \right)+ 2\left(2n_t n_{r}-n_r^{2}-n_r \right) \eta^{2}\Vert\Gmat \Vert^{2}/\delta
\eeq
\end{corollary}
\IEEEproof This is an immediate consequence of   Theorem \ref{TheoremParlett1998Symmetric} in Appendix \ref{ProofOfTheoremQaudraticConvergence}.

Note that the quantity
\(\Vert\Hmat_{12}\Tmat_{k+m}\Vert^{2}\) is the only one that interest the SU that applies the BNSL algorithm.
Furthermore, once \(\eta>> P_{k}^{2}\)   the dominant term in the
   \eqref{BoundOnInterference} will be \(O(\eta)\) i.e., the interference power to the PU will approximately satisfy
\beq
\Vert \Hmat_{12}\Tmat_{m+k}\Vert^{2}\leq 2\left(2n_tn_{r}-n_r^{2}-n_r \right) \eta^{2} \Vert\Gmat\Vert^{2}/\delta
\eeq
This allows  choosing \(\eta\) to guarantee a maximum interference level to the PU, an observation that will be very useful in the simulation part. Theorem
\ref{TheoremQuadraticConvergence} and Corollary \ref{corollary} also imply that the line search accuracy need not be constant during the entire BNSL algorithm but can be refined as the algorithm  goes on.

  The  asymptomatic quadratic convergence rate  of Theorem \ref{TheoremQuadraticConvergence}  and Corollary \ref{corollary} is determined by \(1/\delta\) where \(3\delta\)  is the minimal gap between \(\Gmat\)'s eigenvalues.  In addition, the quadratic convergence rate takes effect only after \(P_{k}^{2}<\delta/8\). Such a condition   implies that if \(\delta\) is very small, it will take the BNSL many cycles to reach its quadratic convergence rate. This is problematic  since MIMO wireless channels may have very close singular values (recall that \(\Hmat_{12}\) square singular values are equal to  \(\Gmat \)'s first \(n_{r}\) eigenvalues). If we were using the optimal  Cyclic Jacobi technique (i.e. no errors due to finite line search accuracy)  this would  not have a practical implications \citep{VjeranSharp1991} since quadratic decrease in \(P_{k}\) that is independent of \(\delta\)  occurs prior to the phase where \(P_{k}^{2}<\delta/8\). In the following theorem we extend this result to the  BNSL algorithm. \begin{theorem}
 \label{TheoremQuadraticConvergenceClusters}
Let $\eta$ be the accuracy of the line search, \(\{\lambda_{l}\}_{l=1 }^{n_{t}}\) be \(\Gmat\)'s eigenvalues such that there exist a cluster of eigenvalues, i.e.   \(\lambda_{l}=\lambda+\xi_{l}, $ for  $ l=n_{t}-v+1,...,n_{t}\) where  \(\sum_{l=n_{ t}-v+1 }^{n_{t}}\xi_{l}=0\). Assume that the rest of the non equal eigenvalues are well separated, i.e. \(\delta_{c}>>\vert \xi_{l}\vert\) where
\beq
  3\delta_{c}=\min(\Lambda_{1}\cup \Lambda_{2})
\eeq
\beq\begin{array}{lll}\Lambda_1=
\left\{
  \vert\lambda_{l}- \lambda_{r} \vert:1\leq l<r\leq n_{t}-v, \; \lambda_{l} \neq \lambda_{r}\right\}\\\Lambda_{2}= \left\{
  \vert\lambda_{l}- \lambda \vert :1\leq l\leq n_{t}-v\right\} \end{array}
  \eeq
Then, once the BNSL algorithm reaches a stage
such that   \(2\delta_{c}\sqrt{ \sum_{l} \xi_{l}^{2}}\leq P^{2}_{k}\leq \delta^{2}_{c}/8\),   and if \(\eta<<P^{2}_{k}\), then
 \begin{equation}
 \label{EtacInequilityOfTheTheorem} \begin{array}{lll}P_{k+m}\leq O\left( \left( \frac{P_{k}} {\sqrt{\delta_{c }}}\right)^{ 2}\right)
\end{array} \end{equation}\end{theorem}
\IEEEproof See Appendix
 \ref{Appendix:ProofOfClusters}.

In the presence of very  close eigenvectors cluster, i.e. \(\sqrt{\sum_{l}\xi_{l}^{2 }}<< \delta_{c}\),  the distance    \(\delta_{c}\) will be    much greater than \(\delta\). In that case,  a quadratic decrease in \(P_{k}\) will occur  even before \(P_{k}\) becomes smaller than  \(\delta
/\sqrt{8}\) but only satisfies     \( 2\delta_{c}\sqrt{ \sum_{l} \xi_{l}^{2}}\leq P^{2}_{k}\leq \delta^{2}_{c}/8\). This quadratic decrease  brakes down and become  slower  as  \(P^{2}_{k}\) becomes smaller  than  \(2\delta_{c}\sqrt{ \sum_{l} \xi_{l}^{2}}\). From a practical point of view, this is not a problem if one is not interested in decreasing \(P_{k}^{2}\)  more than   \({ 2\delta_{c}\sqrt{\sum_{l} \xi_{l}^{2}}}\) which may be very small.  Nevertheless, \(P_{k}\) will  eventually   decrease  quadratically  as \(P^{2}_{k}\) becomes smaller than \(\delta/8\) as required by Theorem
  \ref{TheoremQuadraticConvergence}.    This phenomena is for the Cyclic Jacobi technique  in
\citep{VjeranSharp1991}.

\section{simulations}\label{Section:Simulation}
\begin{figure}
\centering
{ \psfig{figure=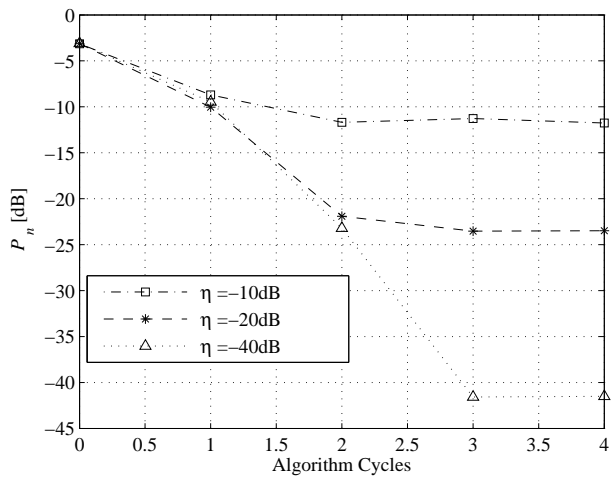, width=9cm}}
\caption{Simulation results for different line search accuracy values \(\eta\)  of the BNSL algorithm for obtaining the null space of \(\Hmat\) with \(n_{t}=3\) transmitting antennas and \(n_{r}=2\) antennas at the PU receiver.
The vertical axis represents the square of the sum  of the magnitude of the upper off-diagonal entries of \(\Gmat=\Hmat^{*} \Hmat\) while the horizontal axis represent the number of  complete  cycles of the BNSL algorithm, i.e. \((n_{t}^{2} -n_{t})/2\) learning phases.  The matrix \(\Gmat \) where normalized such that \(\Vert\Gmat \Vert^{2}=1\). We used 200 Monte-Carlo trails where the entries of  \(\Hmat\)   are i.i.d. complex Gaussian Random variables.}
\label{Figure:Convergence}
\end{figure}

Figure \ref{Figure:Convergence}
presents simulation results of the BNSL algorithm for different levels of line search accuracies. It is shown that for sufficiently small \(\eta\) the   algorithm converges quadratically. The quadratic decrease   breaks down when the  value  of \(P_{k}\) becomes as small as an order of magnitude of    \(\eta\). This result is consistent  with the bound that is proposed in   Theorem \ref{TheoremQuadraticConvergence}. 
\begin{figure}
\centering
{ \psfig{figure=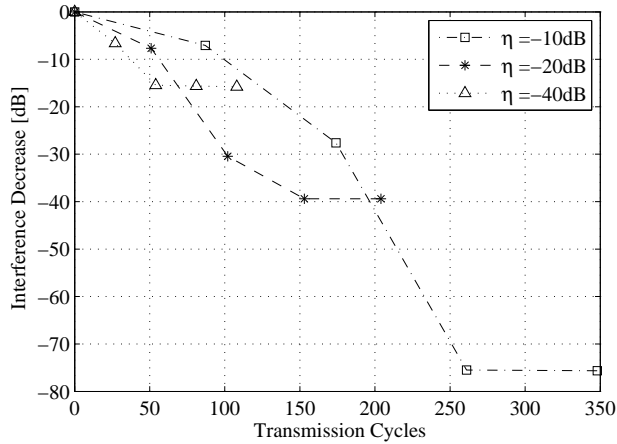, width=9cm}}
\caption{Simulation results for adoptive  line search  accuracy values   of the BNSL algorithm for obtaining the null space of \(\Hmat\) with \(n_{t}=3\) transmitting antennas and different number of  antennas at the PU receiver \(n_{r}\). The vertical axis represents a bound on the norm  of the interference to the PU  while the horizontal axis represent the number of  transmission  cycles.  The matrix \(\Gmat \) where normalized such that \(\Vert\Gmat \Vert^{2}=1\). We used 100 Monte-Carlo trails where the entries of  \(\Hmat\)   are i.i.d. complex Gaussian Random variables.}
\label{Figure:InterferencePlot}
\end{figure}In addition, Figure
 \ref{Figure:Convergence}  shows that   as long as \(\eta\) is smaller or equal to   \( P^{2}_{k}\) the decrease in \(P_{k+m}\) is almost not  affected by  different line search accuracies as demonstrated by the fact that the  decrease is approximately  the same for \(\eta=-10, -20,-40\) at the first cycle  as well as for \(\eta=-20,-40\), at the second cycle. The simulation shows that this phenomena, which is  consistent  with the asymptomatic behavior as    indicated  by   Theorem \ref{TheoremQuadraticConvergence},  is also valid before the algorithm reaches its asymptomatic behavior. Figure \ref{Figure:InterferencePlot} describe the interference decrease as a function of the transmission cycles for  different line search accuracies. The result is consistent with  Corollary \ref{corollary} where   the ultimate level of the interference is bounded by \(O(\eta^{2})\). Similar to Figure
 \ref{Figure:Convergence} the interference decrease in Figure
 \ref{Figure:InterferencePlot} is approximately  the same for \(\eta=-10, -20,-40\) at the first cycle  as well as for \(\eta=-20,-40\), at the second cycle. Moreover, Figure \ref{Figure:InterferencePlot} shows that increasing the  values of \(\eta\) reduces the number of transmission cycles  drastically as  discussed in Section
  \ref{ReducingTheLineSearchComplexity}.

\begin{figure}
\centering
{ \psfig{figure=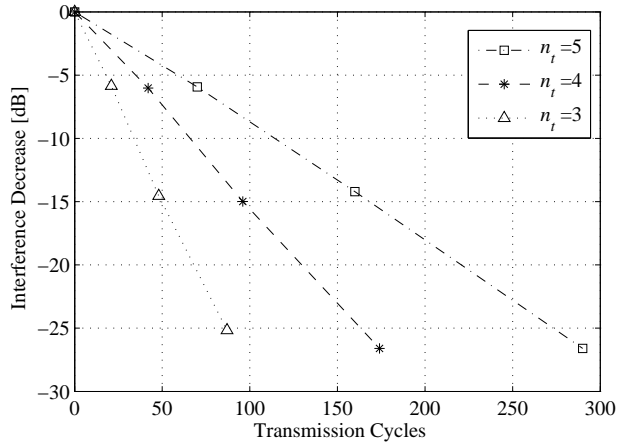, width=9cm}}
\caption{Simulation results    of  BNSL algorithm for non constant   line search accuracy and different numbers of transmitting antenna at the SU transmitter.  The line search accuracy \(\eta\)  is -6dB, -8dB -15dB at the first,  second and  third cycles respectively. The PU has  \(n_{r}=2\) transmitting antennas. The vertical axis represents the   reduction in  the interference to the PU  while the horizontal axis represent the number of  transmission  cycles.   We used 200 Monte-Carlo trails where the entries of  \(\Hmat\)   are i.i.d. complex Gaussian Random variables.}
\label{Figure:InterferencePlotDifferentNt}
\end{figure}

  Both Figures \ref{Figure:Convergence} and  \ref{Figure:InterferencePlot} suggest that    using   a low    line search accuracy (i.e. larger \(\eta\)) in the first  cycle and increasing it from one cycle to the other may reduce the overall transmission cycles with no significant performance loss. This idea is put into practice in Figure \ref{Figure:InterferencePlotDifferentNt} where  we present simulation results of the interference decrease to the PU as a function of the transmission cycles for an increasing line search accuracy.

\section{conclusions}
\label{Section:coclusions}

We proposed a  blind technique for interference mitigation by secondary cognitive users based on a blind implementation of the well known cyclic Jacobi technique. The only condition required is that the primary  user will be using a power control scheme such that for a short time interval, its transmitted power be  effected monotonically by the interference from one  SU. This includes also a case where there are multiple secondary users  in a steady state, i.e. their interference power is constant during this time period. The entire learning process is based on energy  measurements and on detecting energy variations. This means that the secondary users are not required to be synchronized to the PU pulse time. Furthermore the proposed learning scheme is independent of the PU transmission scheme (i.e. coding, modulation) as long as the power allocation is a monotonous function of the interference from the secondary user.

The Convergence  properties of the  BNSL algorithm were   also  explored in this paper. It was shown that  the BNSL algorithm  maintains the convergence  properties of the Cyclic Jacobi technique, that is, a  global linear and an asymptotically quadratic convergence  rates, as long as the line search accuracy is sufficiently small. It was also  shown  in simulation  that just like in the  cyclic Jacobi technique, the BNSL algorithm reaches its  quadratic convergence rate in just  three to four cycles. Furthermore, we obtained a bound on the interference that the SU inflicts to the PU as a function of the line search accuracy of the BNSL algorithm and  provided a mechanism for choosing this line search accuracy to reduce the number of transmission cycles while maintaining low levels of interference to the PU.

It is important to stress that the BNSL algorithm is not necessarily limited to  energy measurements taken by the SU  and to PU that apply power control.  The BNSL algorithm can also  be implemented in other scenarios as long as the SU can learn whether the interference it inflicts to the PU has  increased or decreased between one transmission cycle to the other. For example, the secondary user can learn about the interference  by observing  the PU's modulation scheme since this is too a function of the the interference at the PU. An alternative way for implementing the  BNSL algorithm is for the SU to listen to the PU's control channel and extract information about the interference it inflict to the PU.

\appendices
\section{Proof of Theorems
\ref{Theorem:blind step Theorem} and
\ref{Theorem:TweDimensionalSearchToOneDemenssionalSearch}}
\label{AppendixProofOfTheBlindStepTheorem}
The idea behind the proof  is that each  \(l,m\)
rotation is equivalent to diagonalizing a \(2\times 2\) matrix \(\tilde \Gmat_{
l,m}\) using the rotation matrix \(\tilde\Rmat (\theta, \phi)\) as follows

\beq\begin{array}{ccc}
\tilde\Gmat_{l,m}=\begin{bmatrix} g_{l,l} & g_{l,m} \\
g_{m,l} & g_{m,m} \\
\end{bmatrix},&\tilde\Rmat(\theta, \phi)=\begin{bmatrix}\cos(\theta) & e^{-i\phi}\sin(\theta ) \\
-e^{i\phi}\sin(\theta ) & \cos(\theta) \\
\end{bmatrix}\end{array}
\eeq
Let  $\tilde\rvec_{1}(\theta, \phi),
\tilde \rvec_{2}(\theta,\phi)$ be $ \tilde\Rmat( \theta,\phi)$'s first and second columns respectively. It follows that
 \beq
 \begin{array}{lll}
 [\Rmat_{l,m} (\theta,\phi) \Gmat
 \Rmat_{l,m}^* (\theta,\phi)] _{l,l}
 =\rvec_{l,m}^* (\theta,\phi) \Gmat\rvec_{
 l,m}(\theta,\phi)=
\tilde \rvec_{1}^*(\theta,\phi)\tilde \Gmat_{l,m}\tilde\rvec_{1}(\theta, \phi) \end{array} \eeq

The first part of the Theorem
\ref{Theorem:blind step Theorem} follows directly from the   Rayleigh-Ritz Theorem \citep[see e.g.][Theorem 4.2.2]{horn_and_johonson} that asserts that \beq
\label{rayleigh-ritz}.
\lambda^{\rm min}_{l,m}=\min_{\xvec \in \tilde B}\xvec^* \tilde\Gmat_{l,m} \xvec
\eeq
where \(\tilde B=\{\xvec\in {\mathbb C^{2}} :\Vert\xvec\Vert=1\}\) and $\lambda_{l,m}^{\rm min}$ is $\tilde \Gmat_{l,m}$'s  minimal eigenvalue. It follows that \begin{equation}
 \label{transformingRRTheorem} \lambda_{ l,m}^{\rm min}=\tilde\rvec_{1}^*(\hat \theta, \hat\phi) \tilde\Gmat_{l,m} \tilde\rvec _{1}(\hat\theta,
\hat \phi)\end{equation} where  \beq (\hat\theta,\hat \phi)=\mathop { \arg \min }\limits_{\tiny \theta  ,\phi  \in {\mathbb R}} \tilde\rvec_{1}^*( \theta, \phi) \tilde\Gmat_ {l,m} \tilde\rvec_{1}( \theta, \phi)
\eeq
  Equality  \eqref{transformingRRTheorem} is due to the fact that for  every   $\xvec\in \tilde B$ there exist \(\theta,\phi\in \real\)  such that  $\tilde\rvec_ {1}(\theta,\phi)=a\xvec$  where,\(\vert a\vert=1,a\in\comp\) and because for every \(\Gmat\in\comp^{2\times 2}\) and \(  \xvec\in \comp^{2}\)   the function \(S\) satisfies  \(S(\Gmat ,a\xvec) =S(\Gmat,\xvec)\)  it follows that $\max_{x\in \tilde B}S(\tilde\Gmat_{l,m} ,\xvec)\leq \max_{\theta,\phi\in\real} S(\tilde\Gmat_{l ,m},\tilde \rvec_1(\theta, \phi))$. In addition, because \(\tilde \rvec_{1}(\theta, \phi)\in \tilde B\) for every \(\theta,\phi\) we have that $\max_{x\in \tilde B} S(\tilde\Gmat_{ l,m},\xvec) $ $\geq $ $ \max_{\theta, \phi\in\real} S(\tilde\Gmat_{l,m}, $ $\tilde \rvec_1(\theta, \phi))$ which establishes \eqref{transformingRRTheorem}.
      Note  that $\tilde\rvec_{1}(\hat  \theta, \hat\phi)$ is the eigenvector that corresponds  $\lambda_{l,m}^{\rm min}$  and since $\tilde\rvec_{1}$ and $\tilde\rvec_{2}$ are orthonormal, it follows that $\tilde\rvec_{2}$ is the eigenvector that corresponds to $\tilde \Gmat_{l,m}$'s maximal eigenvalue $\lambda_{l,m}^{\rm max}$. Hence,
\beq
\tilde\Rmat (\hat \theta,\hat \phi)\tilde \Gmat_{l,m}\tilde\Rmat^* (\hat \theta, \hat \phi) =\begin{bmatrix} \lambda_{l,m }^{\min} & 0 \\
0 & \lambda_{l,m}^{\max} \\
\end{bmatrix}
\eeq
and because $[\Rmat_{l,m} (\hat\theta, \hat\phi) \Gmat\Rmat_{l,m}^* (\hat\theta ,\hat\phi)] _{l,m}=[\tilde \Rmat (\hat \theta,\hat \phi)\tilde \Gmat_{l,m} \tilde\Rmat^* (\hat \theta,\hat \phi) ]_{1,2}=0 $, the desired result follows.

It remains to show the second part of the theorem.
The objective function for minimization is
\beq\label{objective}\begin{array}{lll} S\left(\Gmat ,\rvec_{l,m}(\theta, \phi)\right) =[\Rmat_{l,m} (\theta,\phi) \Gmat\Rmat_{l,m}^* (\theta,\phi)] _{l,l}=\cos ^2(\theta ) \left|g_{l,l} \right| \\~~~~~~~~~~~~~~~~~~~~~~~~+\sin ^2(\theta ) \left|g_{m,m}\right|- \left|g_{l,m}\right| \sin (2 \theta ) \cos ( \phi+\angle g_{l,m} )\end{array} \eeq
which is  a continuous and differentiable (of any order) function of $\theta$ and $\phi$. Recall that $\Gmat \geq0$, thus $g_{ll}$ are real non-negative numbers and  therefore will be  written as  $g_{ll}$ instead of $\vert g_{ll} \vert$. We first  assume that \(g_{12}\neq 0\). Setting the gradient to zero yields

\beq
\begin{array}{lll}
0=-2 g_{1,2} \cos (2 \theta ) \cos \left(\angle g_{l,m}+\phi \right)+g_{1,1} (-\sin (2 \theta ))+g_{2,2} \sin (2 \theta )\\0=g_{1,2} \sin (2 \theta ) \sin \left(\angle g_{l,m}+\phi \right)
\end{array}
\eeq and the solution is  \begin{equation} (\theta,\phi)\in\left\{\left((-1)^{a} \theta_{0 }+\frac {b\pi } 2,a\pi-\angle g_{l,m} \right ),\left({\frac{{\pi b}}{2}},{ - { \gamma _{12}} + \frac{\pi }{2} + \pi a}\right) \right\}_{a,b\in {\mathbb Z}}  \end{equation} where
\beqna\label{thetaSol}\theta_{0}=\left\{ \begin{array}{lll} \frac{1}{2} \tan ^{-1}\left( \frac{-2\vert g_{l,m}\vert} {  g_{l,l}- g_{m,m} }\right) &{\rm if }&\; g_{ll}\neq g_{mm}\\ \frac \pi 4 &{\rm if}  &g_{ll}=g_{mm}\end{array} \right.
 \eeqna
We begin with the family of suspected points  \(((-1)^{a}\theta_{0}+\frac {b\pi } 2,a\pi-\angle g_{l,m})\). Since \(S(\Gmat,\rvec_{1}(\theta,\phi) )=S(\Gmat,\rvec_{1}(\theta+\pi,\phi+2 \pi)),\; \forall \theta,\phi\in\real\) it is sufficient to investigate the following subset
\beqna\label{SuspectedPoints}
\begin{array}{lll}(\theta,\phi) \in\{ \gammavec_{1}=(\theta_ {0},-\angle g_{l,m}),\gammavec_{2}=(\theta_{0}+\pi/2,- \angle g_{l,m})\\ ~~~~~~~~~~~~,\gammavec_{3}=( -\theta_ {0},\pi-\angle g_{l,m}),\gammavec_{ 4}( \pi/2-\theta_{0},\pi-\angle g_{l,m}\} \end{array}
\eeqna
Furthermore, because  \(\rvec_{1}( \gammavec_{1} )=-\rvec_{1}(\gammavec_{3})\) and \(\rvec_{1}(\gammavec_{2})=-\rvec_{1} (\gammavec_{4})\) it is sufficient to check the  points  \(\gammavec_{1}\) and \(\gammavec_{2}\).
To investigate these points we calculate
the Hassian of $S(\Gmat,\rvec_{1}(\theta, \phi))$ that is given by
\beq
\begin{array}{lll}
\bigtriangledown^{2}(S)_{\theta, \theta}=4 \sin (2 \theta ) \left|g_{l,m} \right| \cos \left(\phi +\angle  g_{l,m} \right)+2 \cos (2 \theta ) \left(g_{m,m} -g_{l,l}\right)\\
\bigtriangledown^{2}(S)_{\theta, \phi}=2 \cos (2 \theta ) \left|g_{l,m} \right| \sin \left(\phi +\angle  g_{l,m} \right)\\
\bigtriangledown^{2}(S)_{\phi ,\phi}=\sin (2 \theta ) \left| g_{l,m}\right| \cos\left(\phi +\angle  g_{l,m}\right)
\end{array}
\eeq
and
its primary minor (determinant)  \beq\label{HessianDet} \begin{array}{lll}
\det\left(\bigtriangledown^{2}(S) \right)= \left|g_{l,m}\right| \left(2 \left|g_{l,m}\right| \left(\cos \left(2 \phi +2 \angle  g_{l,m}\right)-\cos (4 \theta )\right)\right.\\ ~~~~~~~~~~~~ ~~~~~~~~~~\left.+ \sin (4 \theta ) \left(g_{m,m}- g_{l,l}\right) \cos \left(\phi +\angle  g_{l,m}\right)\right)
\end{array}
\eeq
is equal  to \(4\vert g_{l,m}\vert^{2}\)    for \(\gamma_{j},j=1,...,4\). Therefore these are  local extreme points.  In order to determine which is the maximum and which is the minimum we will substitute these points in $\bigtriangledown^{2 }(S)_{\theta,\theta}$
\beqna\label{FirstPoint}
\bigtriangledown^{2}(S)_{\theta,\theta} (\theta_{0},-\angle g_{l,m})&=& \left\{ \begin{array}{lll}-\frac{\vert\cos (\theta_{0})\vert2 \left(4 g_{1,2}^2+\left( g_{1,1}-g_{2,2}\right) {}^2 \right)}{ \left( g_{1,1}-g_{2,2}\right) }&{\rm if}\; g_{11}\neq g_{22}\\ 4g_{1,2}& \text{otherwise}\end{array} \right.
\\ \label{SecondPoint}
\bigtriangledown^{2}(S)_{\theta, \theta}( \theta_{0}+\pi/2,-\angle g_{l,m}) &=&\left\{\begin{array}{lll} \frac{2  \left(g_{1,1}-g_{2,2}\right)} {\vert \cos\theta_{0}\vert}&{\rm if}\; g_{11} \neq g_{22}\\ -4  g_{12}& \text{otherwise} \end{array}\right.
\eeqna
Note that \eqref{FirstPoint} and \eqref{SecondPoint} are of opposite signs, therefore,   one  of which corresponds to a local minimum  while the other to a local maximum. It remains to check the  family of suspected points \(({\frac{{\pi b}}{2}},{ - {\gamma _{12}} + \frac{\pi }{2} + \pi a})\). By substituting it in  \eqref{HessianDet} we see that it is equal to -\(4\vert g_{l,m}\vert^{2}\) for every \(a,b\in {\mathbb Z}\). It   follows that these suspected points are not extreme points.
This establishes that for \(g_{1,2}\neq 0\) all of the  of   local minimum points (which are   infinitely countable) of \(S(\Gmat,\rvec_{1}(\theta,\phi))\) are equivalent, i.e. for every local minimum points \((\hat\theta,\hat \phi), (\theta',\phi')\) such that  \((\hat\theta,\hat \phi)\neq (\theta',\phi')\) we have \(S(\Gmat,\rvec_{1}(\hat\theta,\hat \phi))=S(\Gmat,\rvec_{1}(\theta',\phi'))\). Thus, every local minimum is also a global minimum. In the case where \(g_{1,2}=0\), the  target function is
\beq
S\left(\Gmat ,\rvec_{l,m}(\theta, \phi)\right) =\cos ^2(\theta )  \left|g_{ l,l}\right| +\sin ^2(\theta ) \left|g_{m,m }\right|
\eeq
which obviously satisfies the conditions of Theorem \ref{Theorem:blind step Theorem}. This establishes the proof of Theorem \ref{Theorem:blind step Theorem}.

To proof Theorem
 \ref{Theorem:TweDimensionalSearchToOneDemenssionalSearch} we  substitute  $\theta=\pi/3$ in
\eqref{objective}
  \beq\label{s of 0.5}
S\left(\Gmat ,\rvec_{l,m}(\pi/3, \phi) \right) =-\frac{1}{2} \sqrt{3} \left|g_{l,m} \right| \cos \left(\phi +\angle  g_{l,m}\right)+ \frac{3 \left|g _{l,l} \right| }{4}+\frac{\left |g_{m,m}\right|}{4}
\eeq  It follows that in  the blind one-dimensional minimization  $\min_{\phi} S\left(\Gmat ,\rvec_{l,m}(\pi/3, \phi)\right)$,   $\cos(\tilde\phi+ \angle g_{l,m})=1$ which is obtained by $\tilde\phi =-\angle g_{l,m}$ since the line search is carried out  in the interval \(\phi\in[-\pi,\pi]\). By  performing blindly the optimization in  \eqref{tilde phi Optimization}  we obtain the minimum that corresponds to either \(\gammavec_{1}\) or \(\gammavec_{ 2}\) in \eqref{SuspectedPoints}. Then, the value of $\hat\theta$ is chosen  such that \(\hat\theta\)  lies in the interval \([-\pi/4,\pi/4]\), thus \((\hat \theta,\hat \phi)\in \{\gamma_{1},\gamma_{3}\}\) where \(\gamma_{1},\gamma_{3}\) are defined in \eqref{SuspectedPoints}.
 \hfill $\Box$

\section{Proof of Theorem \ref{Proposition Linear Convergence}}
\label{Appendix3}
Consider the first cycle of  the cyclic Jacobi technique, i.e.  $k=1,2,..., n_{t}(n_{t}-1)/2$.   Denote the number of rotated elements in the $l$th row by  \(b_{l}=n_{t}-l\) and let \beq
\begin{array}{lll} c_{l} =\sum_{j=1}^{l} b_{j}=(2n_{t}-1-l)l/2\\ Z(l,k)= \sum_{j=1 }^{n_{t}-l}\vert [\Amat_{ k}]_{l,j+l}\vert^{2} \\
W(l,k)=\sum_{j=l+1}^{n_{t}-1}  Z(j,k)
\end{array}\eeq
Note that \(W(0,k)=P^{2}_{k}\).
In every cycle, each entry is eliminated once, we therefore  denote \(\Amat_{k}\)'s $p,q$ entry  before its annihilation  as  $g_{q,p}(t)$ where $t$ denotes the number of changes since  $k=0$. After $g_{q,p}(t)$ is annihilated, it will be denoted by $\tilde g_{q,p}(\tilde t)$ where $\tilde t$ is the number of changes after the annihilation. The   diagonal entries of $\Amat_{k}$ will be denoted by $x$   since we are  not interested in their values in the course of the proof. This is illustrated in the following example of a $4\times 4$ matrix
\beq\begin{array}{cc}
\begin{array}{c}
\Amat_0=\Gmat
 \\
\left(
\begin{array}{cccc}
 g_{1,1}(0) & g_{1,2}(0) & g_{1,3}(0) & g_{1,4}(0) \\
 g_{2,1}(0) & g_{2,2}(0) & g_{2,3}(0) & g_{2,4}(0) \\
 g_{3,1}(0) & g_{3,2}(0) & g_{3,3}(0) & g_{3,4}(0) \\
 g_{4,1}(0) & g_{4,2}(0) & g_{4,3}(0) & g_{4,4}(0)
\end{array}
\right) \\
\end{array} & \begin{array}{c}
\Amat_1 \\
\left(
\begin{array}{cccc}
 x_{} & \epsilon  & g_{1,3}(1) & g_{1,4}(1) \\
 \epsilon  & x_{} & g_{2,3}(1) & g_{2,4}(1) \\
 g_{3,1}(1) & g_{3,2}(1) & x_{} & g_{3,4}(0) \\
 g_{4,1}(1) & g_{4,2}(1) & g_{4,3}(0) & x_{}
\end{array}
\right) \\
\end{array} \\
\begin{array}{c}
\Amat_2  \\
\left(
\begin{array}{cccc}
 x & \tilde{g}_{1,2}(0) & \epsilon  & g_{1,4}(2) \\
 \tilde{g}_{2,1}(0) & x & g_{2,3 }(2) & g_{2,4}(1) \\
 \epsilon  & g_{3,2}(2) & x & g_{ 3,4}(1) \\
 g_{4,1}(2) & g_{4,2}(1) & g_{4, 3}(1) & x
\end{array}
\right) \\
\end{array} & \begin{array}{c}
\Amat_3 \\
\left(
\begin{array}{cccc}
 x & \tilde{g}_{1,2}(1) & \tilde{g}_{1 ,3}(0) & \epsilon  \\
 \tilde{g}_{2,1}(1) & x & g_{2,3 }(2) & g_{2,4}(2) \\
 \tilde{g}_{3,1}(0) & g_{3,2}(2) & x & g_{3,4}(2) \\
 \epsilon  & g_{4,2}(2) & g_{4,3}(2) & x
\end{array}
\right) \\
\end{array} \\
\end{array}
\eeq
For arbitrary \(n_t\), after the first $c_{1}$ sweeps $\Amat_{c_{1}}$'s  first column is equal to the following vector:
\beq
[x,\tilde g_{2,1}(n_{t}-3),...,\tilde g_{n_{t}-1,1}(0),\epsilon]^{\rm T}
\eeq
and \beq Z(1,c_{1})\leq\vert \tilde g_{2,1}(n_{t}-3) \vert^{2}+...+\vert \tilde g_{n_{t}-1,1}(0)\vert^{2}+\vert \epsilon \vert^{2}
\eeq
For $q=2,...,n_{t}$ we have
\beq
\begin{array}{rll}
\tilde g_{q,1}(n_{t}-q-1)&\leq&\cos \left(\theta _{n_{t}-1}\right) \tilde{g}_{q,1}(n-q-2)-e^{i \phi _{n_{t}-1}} g_{q,n_{t}}(1) \sin \left(\theta _{n_{t}-1}\right)\\ ~&\vdots&~ \\\tilde g_{q,1}(1)&\leq &\cos \left(\theta _{q+1}\right) \tilde{g}_{q,1}(0)-e^{i \phi _{3}} g_{q,q+2}(1) \sin \left(\theta _{3}\right)
\\
\tilde g_{q,1}(0)&\leq&\vert\epsilon  \vert\cos \left(\theta _q\right)-e^{i \phi _q} g_{q,q+1} (1)\sin \left(\theta q\right)
\end{array}
\eeq
 where $\tilde g_{q,1}(-1)=\epsilon$.   Bounds on $\{ \tilde g_{q,1}(l)\}_{l=0}^{n_{t}-q-1}$ can be obtained recursively (i.e.  obtaining bound on  $\tilde g_{q,1}(0),$ substituting and obtaining bound on $\tilde g_{q,1}(1)$ and so on...)
\beq\begin{array}{lll}\label{tilde g}
\tilde{g}_{q,1}(n_{t}-q-1)\leq\epsilon  \prod _{v=q}^{n_{t}-1} \cos \left(\theta _v\right)-\sum _{j=q}^{n_{t}-1} e^{i \phi _j} \sin \left(\theta _j\right) g_{q,j+1}(1) \prod _{v=j+1}^{n_{t}-1} \cos \left(\theta _v\right)\\~~~~~~~ ~~~~~~=\vvec(q)^{\rm T}\zvec(q)+ \vert\epsilon\vert \prod _{v=q}^{n_{t}-1} \cos \left(\theta _v\right)\end{array}
\eeq where  $\vvec,\yvec\in\mathbb{C}^{ n_{t}-q}$ such that  $[\vvec(q)]_{j}=-e^{i \phi _{j+1}} g_{q,j+q}(1)$ and $[\yvec( q)]_{j}=\sin \left(\theta _{j+q-1}\right) \prod _{v=j+q}^{n_{t}-1} \cos \left(\theta _v\right)$.  It follows that
\beq\label{before proposition8}
\begin{array}{lll}
\vert \tilde{g}_{q,1}(n_{t}-q-1) \vert^{2} &\leq\vert\yvec^{\rm T}(q)\vvec(q)\vert^{ 2}+\vert \epsilon\vert^{2} \prod _{v=q}^{n_{t}-1} \cos^{2} \left(\theta _v\right) \\ &\leq \Vert\yvec(q)\Vert^{ 2}\Vert \vvec(q) \Vert^{2}+\vert \epsilon\vert^{2} \prod _{v=q}^{n_{t}-1} \cos^{2} \left( \theta _v\right)
\end{array}\eeq

\begin{proposition}
\label{proposition8}
\beq
\Vert\yvec(q)\Vert^{2}=1-\prod_{i=q }^{n-1} \cos(\theta_{i})
\eeq
\IEEEproof
This is shown by induction. Note that \beq\Vert \yvec(q)\Vert^{2}= \sum_{i=q}^{ n-1} \sin^{2}(\theta_{i}) \prod_{v=i+1}^{n-1}\cos^{ 2}(\theta_{v })\eeq where  \beq \label{emptyProd} \prod_{i=l}^{m }c_{i} \triangleq1, \rm{if} ~l>m.\eeq Assume that
\beq\label{induction assumption}
1-\prod_{i=q}^{m-1}\cos^{2}(\theta_{i})= \sum_{i=q }^{m-1}\sin^{2}(\theta_{i}) \prod_{v=i+1}^{m- 1}\cos^{2}(\theta_{v})
\eeq
is true for some $m\in {\mathbb N}$.
Then, for $m+1$ we have
\beq\begin{array}{lll}
\sum_{i=q}^{m}\sin^{2}(\theta_{i}) \prod_{v=i+1}^{m}\cos^{2}(\theta_{v})\\~~~~~ =\sum_{ i=q}^{m-1} \sin^{2}(\theta_{i}) \prod_{v=i+1 }^{m}\cos^{ 2}(\theta_{v})+\sin^{ 2}(\theta_{m}) \prod_{v=m+1}^{m}\cos^{2}( \theta_{v})\\~~~~~= \cos^{2}( \theta_{m} )\sum_{i=q}^{m-1} \sin^{2}(\theta_{i}) \prod_{v=i+1}^{m-1} \cos^{2} (\theta_{v })+\sin^{2}(\theta_{m})
\end{array}
\eeq
where the last equality is due \eqref{emptyProd}. According to the supposition \eqref{induction assumption} \beq
\begin{array}{lll}
\cos^{2}(\theta_{m})\left( 1-\prod_{i=q}^{ m-1} \cos^{2}(\theta_{i}) \right)+\sin^{2}( \theta_{m})\\
 ~~~~~~~~~~=1-\prod_{i=q}^{m}\cos^{ 2}(\theta_{i})
\end{array}
\eeq
\end{proposition}
  \hfill $\Box$

  By substituting  Proposition
  \ref{proposition8}, into \eqref{before proposition8}
 one obtains\beq\label{after proposition8}
\begin{array}{lll}
\vert \tilde{g}_{q,1}(n_{t}-q-1) \vert^{2} \leq \underbrace{\left(\sum_{i=1}^{ n_{t}-q} \vert g_{q,i+q}(1)\vert^{2}\right) }_{=Z(q,c_{1})}  \left(1-\prod_{i= c_{0+}q}^{n_{t}-1}\cos^{2}(\theta_{i}) \right)+\vert \epsilon\vert^{2} \underbrace{ \prod _{v=q}^{n_{t}-1} \cos^{2} \left(\theta _v\right)}_{\leq 1}
\end{array}\eeq
thus,
\beq\label{29_5_11_1}
\vert \tilde{g}_{q,1}(n_{t}-q-1) \vert^{2}\leq \left(1-\prod_{i=c_{0}+q}^{n_{t}-1} \cos^{2}( \theta_{i})\right)Z(q,c_{1})+ \vert\epsilon \vert^{2}
\eeq
and by summing both sides of \eqref{29_5_11_1} over $q=2,...,n_{t}$  \begin{equation} \label{First Z}\begin{array}{lll}
Z(1,c_{1})\leq\sum_{q=2}^{n_{t}}\left(1-\prod_{ i=c_{0}+q}^{n_{t}-1} \cos^{2}(\theta_{i}) \right)Z(q,c_{1})+(n_{t}-1)\vert\epsilon \vert^{2}\\ ~~~~~\leq(1-\prod _{i=c_0+2 }^{c_1} \cos ^2( {\theta }_i))\underbrace{ \sum_{q=2}^{n}Z(q,c_{1})}_ {W(1,c_{1}) }+(n_{t}-1)\vert\epsilon\vert^{2}\\
~~~~\leq(1-\prod _{i=c_0+2}^{c_1} \cos ^2( {\theta }_i))W(0,0)+(n_{t}-1) \vert\epsilon\vert^{2}
\end{array}
\end{equation}
where the last inequality is due to   $P_{c_{1}}=W(1,c_{1})+Z(1,c_{1}), W(0,0)=P_{0}$, and because $P_{k}$ is a monotonically decreasing sequence \footnote{Forsythe and  Henrici \citep{ forsythe1960cyclic} showed that  the sequence $P_{k}$ is a monotonically decreasing sequence.}. It follows that \begin{equation}\label{z approximation}
Z(1,c_{1})=\sin ^2\left(\Psi _{c_{0}+2 ,c_1}\right) W(0,0)+(n_{t}-1)\vert\epsilon \vert^{2}
\end{equation}
where
\beq \label{definition of psi}
\sin ^2\left(\Psi _{c_{l-1}+2,c_l} \right)=1-\prod _{i=c_{l-1}+2}^{c_l} \cos ^2( \tilde{\theta }_i)
\eeq
and $\vert\tilde \theta_{i}\vert \leq\vert \theta_{i}\vert $.
Thus, \begin{equation}
P_{c_{1}}=W(1,c_{1})+Z(1,c_{1}) \leq W(0,0)=P_{0}
\end{equation}
substituting \eqref{z approximation}
we obtain

\begin{equation}
  W(1,c_{1})\leq W(0,0) \cos ^2\left( \Psi _{2,c_1}\right)-(n_{t}-1)\vert \epsilon \vert^{2}
\end{equation}

Now that this relation is  established, it will be applied to \(\Amat_{c_1}\)'s lower \((n_{t}-1)\times (n_{t}-1)\) block-diagonal.  To do that we use the fact that for $J_{c_{1}+1} =(2,3),...,J_{c_{2}}=(2,n_{t}-1)$  the sum of squares of the first column (which is equal to the first row) remains unchanged, i.e. $\sum _{l=1 }^{n_{t}}\vert [\Amat_{k_{}}]_{1,n}\vert^{ 2}$ is constant for $k=c_{1},c_{1}+1,... ,c_{2}$. Thus
\begin{equation}
 W(l,c_{l})\leq W(l-1,c_{l-1}) \cos ^2\left(\Psi _{c_{l-1}+2,c_{l}}\right)-(n_{t}-l)\vert \epsilon\vert^{2}
\end{equation}
Continuing this way we obtain
\beq W(l,c_{l})\leq W(0,0) \prod _{j=1}^l \cos ^2\left(\Psi _{c_{j-1}+2,c_j}\right) -\epsilon ^2 \sum _{j=1}^l b_j \prod _{v=j+1}^l \cos ^2\left(\Psi _{c_{v-1}+2, c_v}\right) \eeq thus \beq \begin{array}{ll}  Z(l,c_{l})=\sin ^2\left(\Psi _{c_{l-1}+2,c_l}\right) W(l-1,c_{l-1})+(n_{t}-l)\vert \epsilon^{2} \vert \\~~~~~~~~~\leq W(0,0) \sin ^2\left(\Psi _{c_{l-1}+2,c_l}\right) \prod _{j=1 }^{l-1} \cos ^2\left(\Psi _{c_{j-1 }+2,c_j}\right)\\~~~~~~~~~~~-\vert\epsilon\vert ^2 \sum _{j=1}^{l-1} b_j \prod _{v=j+1 }^{l-1} \cos ^2\left(\Psi _{c_{v-1}+2,c_v }\right)+(n_{t}-l)\vert \epsilon^{2}\vert
\end{array}\eeq
After a complete cycle
\beq\label{Pcn-1}\begin{array}{lll}
P^{2}_{c_{n_{t}-1}}=&\sum _{l=1}^{n_{t}-2} Z(l,c_{ n_{t}-1})+\vert\epsilon\vert^{2} =\sum _{l=1}^{n_{t}-2} Z(l,c_{l})\\&\leq W(0,0)\sum_{l=1}^{n_{t}-2} \sin ^2\left(\Psi _{c_{l-1}+2,c_l}\right) \prod _{j=1}^{l-1} \cos ^2\left(\Psi _{c_{j-1}+2,c_j}\right) \\&-\sum_{l=1}^{n_{t}-2} \vert\epsilon\vert ^2 \sum _{j=1}^{l-1} b_j \prod _{v=j+1}^{l-1} \cos ^2\left(\Psi _{c_{v-1}+2,c_v}\right)
+\vert \epsilon^{2} \vert\sum_{l=1}^{n_{t}-1 }(n_{t}-l)
\end{array}\eeq
Similar to proposition \ref{proposition8}, it can be shown that \begin{equation} \sum _{l=1}^n \sin ^2\left(\tau _l\right) \prod _{j=1}^{l-1} \cos ^2\left(\tau _j\right)=1-\prod _{j=1}^n \cos ^2 \left( \tau _j\right) \end{equation} Thus
\begin{equation}\begin{array}{lll} P^{2}_{c_{n-1}}\leq& W(0,0)\left( 1-\prod_{ j=1 }^{n-2}\cos ^2\left(\Psi _{c_{j-1}+2,c_j }\right)\right)\\&-\sum_{l=1}^{n-2} \epsilon ^2 \sum _{j=1}^{l-1} b_j \prod _{v=j+1}^{l-1} \cos ^2\left(\Psi _{c_{v-1}+2,c_v}\right)
+\vert \epsilon^{2} \vert\sum_{l=1}^{ n-1} b_{l} \end{array}\end{equation}
From \eqref{definition of psi} we have
\begin{equation}\cos ^2\left(\Psi _{c_{l-1}+2,c_l}\right)\geq\prod _{v=c_{l-1}+2}^{ck_l} \cos ^2( \theta _{ v}) \end{equation} and therefore \begin{equation}\begin{array}{lll} P^{2}_{c_{ n-1}}\leq& W(0,0)\left( 1-\prod_{ j=1}^{n-2}\prod _{v=c_{j-1}+2}^{c_j} \cos ^2(\theta_{v})\right)\\&-\sum _{l=1}^{n-2} \vert\epsilon\vert ^2 \sum _{j=1}^{l-1} (n-j) \prod _{v=j+1 }^{l-1} \prod _{r=c_{v-1}+2}^{c_v} \cos ^2\left(\theta _r\right)+ \frac{ \vert \epsilon^{2}\vert(n^2 - n)}{2}  \end{array}\end{equation}
Recall that $\vert\theta_{i}\vert<\pi/4$, therefore
 \begin{equation}\label{EpsolonInequility} \begin{array}{lll}P^{2}_{c_{n-1}}&\leq&  W(0,0)\left( 1-2^{-(n-2) (n-1)/2 } \right) \\ & &-\vert\epsilon ^2\vert \left(\sum _{l=1}^{n-2}  \sum _{j=1 }^{l-1} (n-j) 2^{\frac{l^2}{2}-l n+\frac{l}{2}+9 n-45}-\frac{ (n^2  -n)}{2} \right) \\ &\leq & W(0)\left( 1-2^{-(n-2) (n-1)/2 } \right)+\vert\epsilon ^2\vert\frac{ (n^2-n)}{2} \end{array} \end{equation}

It remains to relate \(\epsilon\) to the accuracy of the line search  \(\eta\).   Note that the error  \(\epsilon\) in \eqref{EpsolonInequility} results from the two  finite-accuracy (of \(\eta\) accuracy) line-searches
in Table \ref{Blind Jacobi table}
on lines \ref{line:BNSL4} and \ref{line:BNSL6}. If \(\eta\) were zero,   \(\Amat_{k}\)'s \(l,m\)  off diagonal entry would   be  zero  after the \(k\)th sweep, i.e. \beq u(\theta^{\rm opt},  \phi^{\rm opt} )=0\eeq
 where
\beq \begin{array}{lll}
u(\theta,\phi)\define\vert[\Rmat_{l,m} (\theta,\phi) \Amat_{k}\Rmat_{l,m}^* (\theta,\phi)] _{l,m}\vert^{2}=4 (a_{l,m}^k)^{2} \sin ^2\left(\gamma _{l,m}+\phi \right)\\~~~~~~~~~~~+\left(2 \cos (2 \theta ) a^{k}_{l,m} \cos \left(\angle a^{k} _{l,m}+\phi  \right) +\sin (2 \theta ) \left(a^{k}_{l, l}- a^{k}_{ m,m}\right)\right)^2
\end{array}
\eeq
and
\((\theta_{k}^{\rm opt},\phi_{k}^{\rm opt})\) is the  value given in Theorem
\ref{Theorem:TweDimensionalSearchToOneDemenssionalSearch} when substituting $\Gmat=\Amat_k$.
Let \((\theta_{k},\phi_{k})\) be the non optimal value that is obtain  by the two line searches  (on Line \ref{line:BNSL4} and Line  \ref{line:BNSL6}), then \(\vert \epsilon \vert^{2} =\max _{k} u (\theta_{k},\phi_{k})\). The error \(u(\theta_{k},\phi_k)\) can be bounded because \(\phi_{k}^{\rm opt }=\angle a^{k} _{l,m}\), thus \(\phi_{k}=-\angle a^{k} _{l,m}+ \eta_{\phi}\) where \(\vert\eta_{\phi}\vert<\eta\). Thus
\beq\label{FirstdefineU1}
u_{1}(\theta_{k},\phi_{k})=4 (a_{l,m}^k)^{2} \sin ^2\left(\gamma _{l,m}+\phi_{k} \right)\leq 4 (a_{l,m}^k)^{2}\eta^{2}\leq 2\Vert\Gmat\Vert\eta^{2}
\eeq
The second term
\beq \label{SecondefineU2}u_{2}(\theta_{k},\phi_{k})=
\left(2  a^{k}_{l,m} \cos \left(\eta_{\phi}  \right)\cos^{2}(2\theta_{k}) +\sin(2 \theta_{k} ) \left(a^{k}_{l, l}- a^{k}_{ m, m}\right)\right)^2
\eeq
Note  that if \(a_{ll}^{k}=a_{mm}^{k}\), the value \(\theta_{k}=\theta_{k}^{\text{opt}}\in{0,\pi/4}\)  since the line search will not miss these points. Now for the case where \(a_{ll}^{k}\neq a_{mm}^{k}\) we have    $\theta_{k}=\theta_{k}^{s}+\eta_{\theta}$ where
  \beq \label{ThetaKS}\theta_{k}^{s}=\frac{1}{2} \tan ^{-1 }\left( x_{k}\right)\eeq
and \beq\label{DefineX} x_{k}=\frac{2 \vert a^{k}_{l,m}\vert \cos (\eta_{\phi} )}{ a^{k}_{m,m}-a^{k}_{ l,l}}\eeq
Note that
\beq \begin{array}{llll}u_{2}(\theta_{k},\phi_{k})=\left(2 \cos \left(\eta _{\phi }\right) a_{l,m}^k \left(\cos \left(2 \theta _k^s\right)-2 \eta _{\theta } \sin \left(2 \theta ^*\right)\right)
\right. \\\;\;\;\;\;\;\;\;\;\;\;\;\;\;\;\;\;\left.+\left(a_{l,l}^k-a_{m,m}^k\right) \left(\sin \left(2 \theta _k^s\right)+2 \cos \left(2 \theta ^*\right) \eta _{\theta}\right)\right)^2
\end{array}\eeq
where \((\theta^{*},\phi^{*})\) is a point on the line that   connects   the points  \((\theta_{k}^{ \rm opt},\phi^{\rm opt}_{k})\),  \((\theta_{k},\phi_{k})\). By substituting \eqref{ThetaKS} we obtain
\beq \begin{array}{lll} u_{2}(\theta_{k}\phi_{k})=\left(\frac{2 \cos \left({\eta }_{\phi }\right) a_{l,m}^k+x_k a_{l,l}^k-x_k a_{m,m}^k}{\sqrt{x_k^2+1}}- 4 \eta _{\theta } \sin \left(2 \theta ^*\right) \cos \left({\eta }_{\phi }\right) a_{l,m}^k+2 \eta _{\theta } \cos \left(2 \theta ^*\right) \left(a_{l,l}^k-a_{m,m}^k\right)\right)^2 \end{array}
\eeq
  by \eqref{DefineX} and by the fact that sinusoidal is bounded by  one and by \(\vert\eta_{\theta}\vert\leq \eta\) we obtain
\beq\label{BoundU2}\begin{array}{lll}
u_{2}(\theta_{k}\phi_{k})\leq4\eta^{2}\left(2   \vert \sin(2\theta^{*})\vert a_{l,m}^k+  \cos(2\theta^{*})\left\vert a_{l,l}^k-a_{m,m}^k \right\vert\right)^2\\\leq4\eta^{2}\left(4\sin^{2}(2 \theta^{*})\vert a_{l,m}^{k}\vert^{2+} 2\sin(4\theta^{*})\vert a_{l,m}^{k}\vert\vert a_{ll}^{k}-a_{mm}^{k}\vert+\cos^{2}(2\theta^{*})\vert a_{ll}^{k}-a_{mm}^{k}\vert^{2} \right)\\ \leq4\eta^{2}\left(2\vert a_{l,m}^{k}\vert^{2} +2\sin(4\theta^{*})\vert a_{l,m}^{k}\vert\vert a_{ll}^{k}-a_{mm}^{k}\vert+\vert a_{ll}^{k}-a_{mm}^{k}\vert^{2}+2\vert a_{l,m}^{k}\vert^{2} \right) \end{array}\eeq \beq \begin{array}{lll} u_2{(\theta_{k},\phi_{k})}\leq 4\eta^{2}\left(2\Vert \Gmat\Vert^{2}+  \sqrt 2\Vert \Gmat\Vert\Vert \Gmat\Vert+\Vert \Gmat\Vert^{2} \right)\end{array}\eeq
Thus
\beq \label{BoundingEpsilonWithEta}
\begin{array}{lll}
\vert \epsilon \vert ^{2} = \max_{k}u(\theta_{k},\phi_{k})\leq2(7+2\sqrt 2 )\eta^{2}\Vert\Gmat\Vert^{2}
\end{array}\eeq
This expression is substituted into \eqref{EpsolonInequility}
and the desired result follows.
ˆ
\section{Proof of Theorem
\ref{TheoremQuadraticConvergence} }\label{ProofOfTheoremQaudraticConvergence}
We first assume that \(\Gmat\)'s eigenvalues are all distinct. From \eqref{tilde g}  it follows that
\beq\label{before proposition8 monified}
\begin{array}{lll}
\vert \tilde{g}_{q,1}(n_{t}-q-1) \vert^{2} \leq\sum _{j=q}^{n_{t}-1} \sin ^2\left(\theta _j\right) \left|g_{q,j+1}(1)\right|{}^2+ \epsilon ^2 \prod _{v=q}^{n_{t}-1} \cos ^2\left(\theta _v\right)
\end{array}\eeq
Similarly to the derivation of \eqref{29_5_11_1} we have
\beq\label{8_6_11_1}\begin{array}{lll}
\vert \tilde{g}_{q,1}(n_{t}-q-1) \vert^{2} \leq Z(q,c_{1})\sum _{j=q}^{n_{t}-1} \sin ^2\left(\theta _j\right)+ \vert \epsilon\vert^{2}
\\ ~~~~~~~~\leq Z(q,c_{1})\sum _{j=2}^{n_{t}-1} \sin ^2\left(\theta _j\right)+ \vert \epsilon \vert^{2} \end{array}\eeq
and by summing both sides of \eqref{8_6_11_1} (similarly to the derivation of \eqref{First Z}) over $q=2,...,n_{t}$ it follows that  \begin{equation}\begin{array}{lll}
Z(1,c_{1})\leq\left(\sum _{j=2}^{n_{t}-1} \sin ^2\left(\theta _j\right) \right) \underbrace{ \sum_{q=2}^{n_{t}}Z(q,c_{1})}_ {W(1,c_{1})}+(n_{t}-1)\vert\epsilon \vert^{2 }\\
~~~~\leq\left(\sum _{j=2}^{n_{t}-1} \sin ^2\left(\theta _j\right) \right) W(0,0)+(n_{t}-1)\vert\epsilon\vert^{2}
\end{array}
\end{equation}

Now that we established this relation we can apply  it  to the  reduced \(n_{t}-l+1\) lower block diagonal and obtain \beq\begin{array}{ll}  Z(l,c_{l})\leq\left(\sum _{j=c_{l-1 }+1}^{c_{l}} \sin ^2\left(\theta _j\right) \right) W(0)+(n_{t}-l)\vert \epsilon\vert^{2}
\end{array}\eeq
After a complete cycle we have
 \beq\label{CopleteCycle}\begin{array}{lll}
P^{2}_{c_{n_{t}-1}}\leq&\sum _{l=1}^{n_{t}-2} Z(l,c_{n_{t}-1} )+\vert\epsilon\vert^{2} =\sum _{l=1}^{n_{t}-2} Z(l,c_{l})+\vert\epsilon\vert^{2} \\&\leq W(0,0)\sum _{j=1}^{n_{t}(n_{t}-1)/2} \sin ^2(\theta _j)+\vert \epsilon^{2} \vert\sum_{l=1}^{ n_{t}-1}(n _{t}-l)
\end{array}\eeq
We  now  relate \(\sum_{j=1}^{n_{t}(n_{t}-1)/2}\sin ^2(\theta _j)\)  to \(W(0,0)\). Let \(
\lambda_1,...,\lambda_{n_{t}} \) be \(\Gmat\)`s eigenvalues and let \beq
\label{DefineDelta}
3\delta  = \mathop {\min }\limits_{{\rm{i}} \ne j} \left| {\lambda _{l} - {\lambda_{m}}} \right|\eeq and assume that  the algorithm has reached a phase  where
 \beq \label{OffDiagonalAssumption}
 P_{k}^{2}=W( 0,k)<\delta^{2}/8\eeq Note that $|
 {a_{ll}^k - a_{mm}^k} |^{2} = | a_{ll}^k - \lambda _l - a_{mm}^k + \lambda _m +
\lambda _l - \lambda _m |^{2} \geq | \lambda _l - \lambda _m |^{2} - | a^{k}_{ll} - \lambda _l |^{2} - | a^{k}_{mm} - \lambda _m|^{2}$, furthermore,
\citep[][Theorem 1]{HenficSpeed1958} we have \(\vert a^{k}_{ii}-\lambda_{i}\vert \leq\delta/2\). Thus,
\beq \label{101}\vert a^{k}_{ll}-a^{k}_{mm }\vert\geq2\delta-\delta/2-\delta/2=\delta
\eeq
Recall that the optimal rotation   angle satisfies  \(\tan(2\theta_{k}^{ \text{opt}})= 2\vert a^{k}_{l_{k}m_{k}} \vert/\vert a^{k }_{l_{k}l_{k}}- a_{m_{k }m_{k}}\vert\) while the actual  the rotation angel is   \(\theta_{ k}=\theta_{k }^{\text{opt}}+\eta_{\theta}\). It follows that     \beq \label{BoundSinTetak}
\begin{array}{lll}\vert\sin ^{2}(\theta_{k}) \vert\leq \vert\sin^{2}( \theta_{k}^{\text{opt }})\vert +\vert \eta_{\theta} \sin(2\theta_{ k}^{\text{opt}})\vert \leq\frac{1}{4}\vert2 \theta _k\vert^{2}\\ ~~
+\vert \eta_{\theta}\vert \tan(2\theta_{k }^{\text{opt} } ) \vert\leq\frac{1}{2^{2}} \tan^{2}(2\theta_{k}^{\text{opt }})+\vert \eta_{\theta}\vert \tan(2\theta_{ k}^{ \text{opt} } ) \vert
\\~~~\leq\frac{\vert a^{k}_{l_{k},m_{k}} \vert^{2 }}{\delta^{2}}+2\vert\eta_{\theta}\vert \frac{\vert a^{k}_{l_{k},m_{k}}\vert}{\delta}\leq \frac{\vert a^{k}_{l_{k},m_{k}} \vert^{2 }}{\delta^{2}}+ \frac{2\vert\eta_{\theta}\vert\sqrt{ W(0,k)}\vert}{\delta} \end{array}\eeq
It follows that \beq \label{BoundSumSin}\begin{array}{lll}\sum _{k=1}^{n_t(n_t-1)/2} \sin ^2(\theta _k)\leq&\sum _{k=1}^{n_t(n_t-1)/2} \left( \frac{\vert a^{k}_{l_{k},m_{k}} \vert^{2 }}{\delta^{2}}+ \frac{2\vert\eta_{\theta}\vert\sqrt{ W(0,k)}\vert}{\delta} \right)\\& =\frac{1}{\delta^{2}} W(0,k)+\frac{ \eta_{\theta}  (n_{t}^{2}-n_{t})}\delta \sqrt{W(0,k)}
 \end{array} \eeq
By substituting \eqref{BoundSumSin} into \eqref{CopleteCycle} one  obtains
\beq\label{PcBound}
P^{2}_{c_{n_{t}-1}}\leq W(0,0)\left(\frac{1}{ \delta^{2}}W(0,0)+ \frac { (n_{t}^{2}-n_{t})\vert \eta_{\theta}\vert}\delta \sqrt {W(0,k)}\right)+\frac{\vert \epsilon^{2} \vert}{2} \left(n_{t}^2-n_{t}\right),
\eeq

 It remains to relate \(\eta_{\theta}\) to the accuracy of the line search \(\eta\). As a  result of  the error on Line
\ref{line:BNSL4}, \(\eta_{\theta}\) depends  on \(\eta_\phi\) as well. Form the  proof of Theorem \ref{Theorem:blind step Theorem} we know that  if an accurate line search were   invoked, it  would  produce  \(\phi_{k}=-\angle a^{k}_{l,m}\). However, due to the finite accuracy \(\eta\), the line search yields \(\phi_{k}=-\angle a_{lm}^{k} +\eta_{\phi}\), where \(\vert \eta_{\phi}\vert\leq \eta\). Thus, \(\theta_{k}\) is obtained by optimizing a slightly perturbed    version of   \(w(\theta)\), say \(\tilde w(\theta)\), due to the   substitution of   \(\phi_{k}\)   into  \(h_{k}\)    (See Table \ref{Blind Jacobi table} Line \ref{line:BNSL5})
i.e. \beq\label{wOfTheta}
\tilde w(\theta)=S(\Amat_{k},\rvec_{l,m}(\theta,\phi_{k}))=h_{k}(\cos ^2(\theta ) a^{k}_{l,l}-\cos (\eta_{\phi} ) \sin (2 \theta ) a^{k}_{l,m}+\sin ^2(\theta ) a^{k}_{m,m})
\eeq
We first assume that \(a_{ll}^{k}\neq a_{mm}^{k}\).  If both line searches were accurate, the optimal value of \(\theta\) would  be \beq
\theta^{\text{opt}}_{k}=\frac{1}{2} \tan ^{-1}\left(p_{k}\right
)\eeq where \(p_{k}=\frac{2 \vert a^{k}_{l,m}\vert}{ a^{k}_{m,m}-a^{k}_{l,l}}\). If one takes into consideration the non-optimality of the  line-search  on Line \ref{line:BNSL4} and ignores the non-optimality of the line search on  Line \ref{line:BNSL6} then \(\tilde w(\theta)\) is given  in \eqref{wOfTheta} and the optimal value would be  \beq
\theta^{s}_{k}=\frac{1}{2} \tan ^{-1}\left( p_{k}\cos (\eta_{\phi} )\right)\eeq
The difference \(\vert\theta^{\rm opt}_{k}- \theta_{k}^{s}\vert\) is
\beq
\begin{array}{lll}
\vert\theta^{\rm opt}_{k}-\theta_{k}^{s} \vert=\left\vert\frac{1}{2} \tan ^{-1}\left( p_{k}\cos (\eta_{\phi} )\right)-\frac{1}{2} \tan ^{-1}\left( p_{k}\right)\right\vert \\ ~~~~~~~\leq\frac{\vert\eta_{\phi}  \sin \left(\eta_{\phi} ^*\right) p_k\vert}{\cos ^2\left(\eta_{\phi} ^*\right) p_k^2+1} \leq \eta_{\phi}^{2}\frac{\vert p_{k}\vert}{\cos ^{2}(\eta_{\phi})p_{k} ^{2}+1}
\end{array}\eeq
where \(\vert\eta_{\phi}^{*}\vert\leq\eta_\phi\). It can be easily that \beq \frac{\vert p_{k}\vert }{\cos^{2}(\eta_{\phi})p_{k} ^{2}+1}\leq\frac{1 }{\vert \cos(\eta_{\phi})\vert}
\eeq
Because \(\theta_{k}=\theta_{k}^{s}+\eta_{\phi}\) and \(\vert \eta_{\phi}\vert<\eta\),  the accumulated effect of  the finite accuracy on  both Line \ref{line:BNSL4} and Line  \ref{line:BNSL6} is bounded by
 \beq
\eta_{\theta}\leq \eta+ \frac{\eta^{2}}{\vert \cos(\eta)\vert} \eeq
Assuming that \(\eta\) is sufficiently small,  (e.g. \(\eta\leq \pi/20\)) we obtain
 \beq \label{BoundingEtaTheta}
 \eta_{\theta}\leq6\eta/5 \eeq
By substituting \eqref{BoundingEtaTheta} and \eqref{BoundingEpsilonWithEta}  into  \eqref{PcBound}    it follows that

\beq\label{104}\begin{array}{lll}
P^{2}_{c_{n_t-1}}\leq W(0,0)\left(\frac{1}{ \delta^{2}}W(0,0)+\eta \frac {6  (n_{t}^{2}-n_{t})}{5\delta} \sqrt {W(0,0)}\right)\\+(10+2\sqrt 2 )(n_t^{2}-n_t)
\eta^{2}\Vert\Gmat\Vert^{2}
\end{array}\eeq
Thus, as long as \(\eta\) is smaller than \(W(0,0)\), the BNSL will have a quadratic convergence rate  for    \(\Gmat \)  that does not have multiple eigenvalues, i.e. all eigenvalues are distinct. This is not sufficient  since we are interested in  a  matrix \(\Gmat\)  \(n_{t}-n_{r}\) with zero eigenvalues.

To extend the proof to the  case where  the matrix \(\Gmat \) has \(n_{t}-n_{r}\) zero eigenvalues and \(n_{r}\) distinct eigenvalues we shall use the following theorem:
\begin{theorem}[\citep{parlett1998symmetric} \label{TheoremParlett1998Symmetric} Theorem 9.5.1]
Let \(\Amat\) be an \(n_{t}\times n_{t}\) Hermitian matrix  with eigenvalues \(\{\lambda_{l}\}_{l=1}^{n_{t}}\) that satisfy \(\lambda_{1}\neq \lambda_{2}\neq \cdots \neq \lambda_{n_{r}}\neq \)
\(\lambda_{n_{r}+1}=\lambda_{n_{r}+2}= \cdots =\lambda_{n_{t}}=\lambda\). Consider the following partition:
\beq\label{PatritionA}
\Amat=\begin{bmatrix}\Amat_1 & \Bmat  \\
\Bmat & \Amat_2 \\
\end{bmatrix}
\eeq  where \(\Amat_{1}\) is  \(n_{r} \times n_{r}\) and \(\Amat_{2}\) is \((n_{t}-n_{r})\times (n_{t}-n_{r})\) and let \(\delta'>0\). If \(\Vert(\Amat_{1}- \lambda \Imat )^{-1}\Vert< 1/\delta'\), then \beq \Vert \Amat_{2}-\lambda\Imat \Vert\leq  \Vert \Bmat \Vert^{2}/\delta'
\eeq
\end{theorem}

We now apply Theorem
\ref{TheoremParlett1998Symmetric}  to the BNSL Algorithm. Let \(\Amat_{ 1}^{k}, \Amat_{2}^{k},\Bmat^{k}\) be \(\Amat_{k}\)'s  submatrices that correspond to the partition in
\eqref{PatritionA}.
   Recall that in our case, \(\lambda =0\), thus,     \eqref{101}  implies  that \(\Vert \Amat_{1}^{k} \Vert> \delta\). Furthermore, by \citep[Corollary 6.3.4]{horn_and_johonson}, the matrix \(\Amat_{1}^{k}\) is invertible, thus \(\Vert(\Amat_{1}^{k})^{-1}\Vert \leq 1/\delta\), and  form Theorem \ref{TheoremParlett1998Symmetric} it follows that  \beq \label{BoundA2WithB} \Vert\Amat_{2}^{k}\Vert\leq\Vert B_{k}\Vert^{2}/\delta\eeq

   To show how this theorem leads to quadratic convergence we first show that once the BNSL algorithm   reaches a stage  where
 \eqref{OffDiagonalAssumption} is satisfied, the affiliation  of the diagonal entries in the upper \(n_{r}\times n_{r}\) -block remains unchanged, i.e.
if  \(\Amat_{k}\) satisfies \eqref{OffDiagonalAssumption} then   \beq\label{UnChanged} \arg\min_{l\in L}\vert\lambda_{l}-a^{k}_{ll}\vert =\arg\min_{m\in L}\vert \lambda_{m}-a^{k+1 }_{mm}\vert,\;\forall L\subseteq \{1,...,n_{r}\}\eeq
Note that \begin{equation}
\left\vert  a_{l_k,l_k}^k-a_{m_k,m_k}^k \right\vert^{2}\leq\sin ^2\left(\theta _k\right) \left(2 \cos (\theta_{k} ) a_{l_k,m_k}^k \cos \left(\phi _k-\angle{ a_{l_{k},m_{k}}^{k}}\right)+\sin \left(\theta _k\right) \left(a_{l_k, l_k}^k-a_{m_k,m_k}^k\right) \right)^2
\end{equation}
 and that for every \(\theta_{k}\) such that \(1\leq k\leq c_{n_{r}}\)   \eqref{BoundSinTetak} is satisfied. Thus
\beq\begin{array}{lll}
\left\vert  a_{l_k,l_k}^k-a_{ l_k,l_k }^{k+1} \right\vert^{2}\leq&\sin ^2\left(\theta _k\right) \left(a_{l_k,m_k }^k +\sin \left(\theta _k\right) \left(a_{l_k,l_k}^k-a_{m_k,m_k}^k \right)\right){}^2\\&\leq  \sin ^2\left( \theta _k\right) \left(a_{l_k,m_k}^k + \left(\left(a_{l_k,m_k}^k\right)^2/ \delta ^2+2 \eta _{\theta } \left.a_{l_k,m_k }^k\right/\delta \right){}^{1/2} \delta \right){}^2\\&\leq  \sin ^2 \left(\theta _k\right) \left(a_{l_k,m_k}^k + \left(\left(a_{l_k,m_k}^k\right){}^2+2 \delta \eta _{\theta } a_{l_k,m_k }^k\right){}^{1/2} \right){}^2 \\&\leq \text{  }\sin ^2\left(\theta _k\right) \left(a_{l_k,m_k}^k + \left(\left. \delta ^2\right/4+2 \delta \eta _{ \theta } \delta /2\right){}^{1/2} \right){}^2 \\&\leq \text{  }\sin ^2\left(\theta _k\right) \left( a_{l_k, m_k}^k + \delta \left(1/4+ \eta _{ \theta } \right){}^{1/2} \right)^2 \\&\leq  \frac{\delta ^2}{4}\left( 1+4\eta _{\theta }\right) \left(1/2+ \sqrt{1/2+\eta _{ \theta }} \right) ^2\\
\end{array}
\eeq
Note that for \(\eta\leq1/100\) and considering \eqref{BoundingEtaTheta} it follows that
\beq
\vert a_{l_{k},l_{k}}^{k}-a_{l_{k},l_{k }}^{k+1}\vert \leq 0.65 \delta
\eeq
which establishes \eqref{UnChanged}.
Now that \eqref{UnChanged} is established, it follows that for every \(l,m\)  such that \(l\neq m\) and \(  1\leq l\leq n_{r}\), we have  \(\vert a^{k}_{ll}-a^{k}_{mm} \vert\geq \delta\).

After \(c_{n_{t}-r}\) rotations  \eqref{CopleteCycle} can be written as
 \beq
 \label{NotCopleteCycle}
 \begin{array}{lll}
P_{c_{n_{r}}}\leq&\sum _{l=1}^{n_{t}-1} Z(l,c_{n_{r}})+\vert \epsilon^{2} \vert\sum_{l=1}^{n_{r}}(n_{t}-l) \\&\leq W(0,0)\sum _{j=1}^{c_{n_{r}}} \sin ^2(\theta _j)+\sum_{l=n_{r}+1 }^{n_{t}-1}Z(l,c_{ n_{r}})+\vert \epsilon^{2} \vert\sum_{l=1}^{{n_{ r}}}(n_{t}-l)
\end{array}\eeq
Recall that \(\vert \epsilon\vert^{2}\leq max_{k}u(\theta_{k},\phi_{k})\) where \(u_{1}(\theta_{k},\phi_{k})\) and \(u_{2}( \theta_{k},\phi_{k})\) are defined in
 \eqref{FirstdefineU1} and \eqref{SecondefineU2}.
From  \eqref{BoundU2} we have
\beq
u_{2}(\theta_{k},\phi_{k})\leq 4\eta^{2}(4P_{k}^{2}+ 4 P_{k}\Vert\Gmat\Vert+\Vert \Gmat \Vert^{2})
\eeq
Because \(\vert a^{k}_{ll}-a^{k}_{mm} \vert\geq \delta\), \eqref{BoundSumSin} is satisfied and   similarly to \eqref{104} we obtain
\beq
\label{111}
\begin{array}{lll}
P_{c_{n_r}}\leq W(0,0)\left( \frac{1}{ \delta ^{2}} W(0,0)+\eta \frac { (n_{t}^{2}-n_{t})}\delta \sqrt {W(0,k)} \right)\\ +2\left(2n_tn_{r}-n_r^{2}-n_r \right) \eta^{2}(9W(0,0)/2+ 4 \sqrt{W(0,0)}\Vert \Gmat \Vert+\Vert \Gmat \Vert^{2})\\+\sum_{l=n_{r}+1}^{ n_{t}-1}Z(l,c_{n_{r}})
\end{array}
\eeq

It remains to bound the term  \(\sum_{ l=c_{n_{r} +1}}^{n_{t}-1} Z(l,c_{n_{r }})\).   Note that for every \(\theta_{ k}\) such that \(1\leq k\leq c_{n_{r}}\),   \eqref{BoundSinTetak} is satisfied. Let \(Q=\left\{(l,m):1\leq l\leq n_{r}<m\leq n_{t} \right\}\). Because \(a_{ll}^{k}\) and \(a_{mm}^{k}\) are located in \(\Amat_{1}^{k}\) and \(\Amat_{2}^{k}\) respectively, it follows that  for   every \(k\) such that  \((l_k,m_k )\in Q\)   \beq\begin{array}{lll}  \vert a_{q, m_{k}}^{k+1}\vert^{2} \leq \vert a^{k}_{ q,m_{k}}\vert^{2} +\sin^{2}( \theta_{k})\vert a^{k}_{l_{k},q} \vert^{2 }, \;{\rm for }\; n_r<q<m_{k}\\\vert a_{m_{k},q}^{k+1}\vert^{2} \leq \vert a^{k}_{ m_{k},q}\vert^{2} +\sin^{2}( \theta_{k})\vert a^{k}_{l_{k},q} \vert^{2 }, \;{\rm for }\; m_{k}<q\leq n_{t}  \end{array} \eeq
and from \eqref{BoundSinTetak} \beq\begin{array}{lll}\vert a_{m_{k},q}^{ k+1}\vert^{2} \leq \vert a^{k}_{m_{k},q}\vert^{2} +\left( \frac{\vert a^{k}_{l_{k},m_{k} } \vert^{2 }}{\delta^{2}}+2\eta_{ \theta} \frac{ \vert a^{k}_{l_{k},m_{k}}\vert}{ \delta} \right)\vert a^{k}_{l_{k},q} \vert^{2 } \;{\rm for }\; m_{k}<q\leq n_{t}  \\ \vert a_{q,m_{k
}}^{k+1}\vert^{2} \leq \vert a^{k}_{q,m_{ k}}\vert^{2} +\left(\frac{\vert a^{k}_{l_{k} ,m_{k}} \vert^{2 }}{\delta^{2}}+2\eta_{
\theta} \frac{\vert a^{k}_{l_{k},m_{k}}
\vert}{ \delta}\right)\vert a^{k}_{l_{k},q} \vert^{2 }, \;{\rm for }\; n_{r}<q<m_{k }
\end{array}\eeq
These can be bounded by
\beq\begin{array}{lll}
\vert a_{m_{k},q}^{k+1}\vert^{2},\vert a_{
q,m_{k}}^{k+1}\vert^{2}\leq W^{2}(0,0) \left( 1+\frac{1}{\delta^{2} } \right)+ \frac{2\eta_{
\theta}}{\delta}W^{3/2}(0,0)
\end{array}
\eeq

Thus, for every \(k\leq c_{n_{r}}\) \beq\begin{array}{lll}
\sum_{l=n_{r+1}}^{n_{t}-1}Z(l, c_{n_{r}})= \sum \limits_{q=n_{r}+1}^{n_{t}-1} \sum \limits_{t= q+1}^{n_{t}} \vert a^{k}_{q,t}\vert^{2}\leq O \left(\left( \frac {W(0,0)}{\delta} \right)^{2}\right)+O \left(\left( \frac {\eta_{\theta}  W^{3/2}(0,0)}{\delta} \right) \right)
\end{array}\eeq
This, together with \eqref{111} and \eqref{BoundingEtaTheta} show \beq\begin{array}{lll} P^{2}_{c_{n-r}}\leq  O \left(\left( \frac {W(0,0)}{\delta} \right)^{2}\right)+O \left(\left( \frac {\eta  W^{3/2}(0,0)}{\delta} \right) \right) \\
\;\;\;\;\;\;\;\;\;\;\;\;+O \left(\left( \frac {\eta^{2}  W^{1/2}(0,0)}{\delta} \right) \right)+2\left(2n_tn_{r}-n_r^{2}-n_r \right) \eta^{2}\Vert\Gmat\Vert^{2} \end{array}\eeq
Since \(P_{k}\) is a decreasing sequence, the desired result follows.

 \section{Proof of Theorem
 \ref{TheoremQuadraticConvergenceClusters}}
 \label{Appendix:ProofOfClusters}  Let \(\Vmat_{k} \Lambda \Vmat_{k}^{\star}= \Amat_{k}\) be \(\Amat_{k}\)'s EVD, and let \beq  \begin{array}{lll}
\tilde \Amat^{k}=\Vmat_{k}\tilde \bLambda\Vmat^{\star}_{k}\\
\hat \Amat^{k}=\Vmat_{k}\hat\bLambda\Vmat^{ \star}_{k}
  \end{array}\eeq
where
\beq\begin{array}{lll} \tilde \bLambda={\rm diag}(\lambda_{1},\cdots,\lambda _{n_{t}-v-r},\underbrace {0 \cdots 0}_r,\underbrace {\lambda\cdots \lambda}_v )\\ \hat \bLambda={\rm diag}(\underbrace {0\cdots 0}_{n_{t}-v},\xi_{1},\cdots, \xi_{v} )\end{array}  \eeq

Similarly to the derivation of \eqref{101},   there exists a permutation
such that
$\vert a_{ll}^{k}- a_{rr}^{k}\vert\geq
\delta_{c}$ for all $ l,m$ such that   \(l<m\) and $1\leq l\leq n_{t}-v-r$ or \(n_{t}-v-r\leq l\leq n_{t}-v+1\)
and \(l>n_{t}-v\). For such a  permutation, let \(\Amat_{k}\) be partitioned as
 \beq \begin{array}{lll} \Amat^{k} =\begin{bmatrix} \Amat_{11}^{k}  & \Amat_{12}^{k} & \Amat_{13}^{k}\\
\Amat_{21}^{k} & \Amat_{22}^{k} & \Amat_{23}^{k} \\
\Amat_{31}^{k} & \Amat_{32}^{k} & \Amat_{33}^{k} \\
\end{bmatrix} \end{array}\eeq
where \(\Amat^{k}_{22}\in \comp ^{r \times r}\) and \(\Amat^{k}_{33} \in \comp^{v\times v}\).
 Then by \citep[][Lemma 2.3]{VjeranSharp1991}
we have that
\beq
\Vert \Amat_{ll}^{k}\Vert_{\rm  off}\leq \frac{P_{k}^{2}}{2\delta_{c}}, \text{for } l=2,3.
\eeq
where \(\Vert\Amat_{ll}^{k}\Vert_{\rm off}\) represent the sum of squares of \(\Amat_{ll}^{k}\)'s  off diagonal entries.  The rest of the proof is identical the proof of Theorem
\ref{TheoremQuadraticConvergence} from equation \eqref{BoundA2WithB} and forward since \eqref{BoundA2WithB} is satisfied by setting \beq \Amat ^{k}_{1}=\Amat_{1 1}^{k}, ~\Bmat^{ k}=[\Amat_{12}^{k}, \Amat_{13}^{k}], ~\Amat^{k}_{2}=\begin{bmatrix} \Amat_{22 }^{2} & \Amat_{23}^{k} \\
\Amat_{32}^{k} & \Amat_{33}^{k} \\
\end{bmatrix}\eeq

\bibliographystyle{ieeetr}

\end{document}